\documentclass[a4paper,11pt]{article}
\pdfoutput=1 

\usepackage{jheppub} 

\usepackage[T1]{fontenc} 

\usepackage[T1]{fontenc} 
\usepackage[utf8]{inputenc} 
\usepackage{microtype} 
\usepackage{lmodern} 
\usepackage{booktabs} 
\usepackage{mdframed} 
\usepackage{graphicx}
\usepackage{comment}

\usepackage[backgroundcolor=white]{todonotes}

\usepackage{hyperref}
 \usepackage{color}
 \definecolor{dark-red}{rgb}{0.4,0.15,0.15}
 \definecolor{dark-blue}{rgb}{0.15,0.15,0.4}
 \definecolor{medium-blue}{rgb}{0,0,0.5}
 \hypersetup{colorlinks, linkcolor={dark-red}, citecolor={dark-blue}, urlcolor={medium-blue}}

\usepackage{amsmath,amsfonts,amssymb}
\usepackage{eucal} 
\usepackage{mleftright} \mleftright
\usepackage{tensor} 
\usepackage{braket} 
\usepackage{bm} 

\usepackage{mathrsfs} 

\providecommand*{\dd}{\mathop{}\!d}
\renewcommand*{\dd}{\mathop{}\!d}

\providecommand*{\pd}{\partial}
\renewcommand*{\pd}{\partial}
\providecommand*{\vd}{\mathop{}\!\delta}
\renewcommand*{\vd}{\mathop{}\!\delta}

\providecommand*{\cc}{\mathsf{\Lambda}}
\renewcommand*{\cc}{\mathsf{\Lambda}}

\title{\boldmath Fractons on curved spacetime in $2+1$ dimensions}

\author[a,1]{Jelle Hartong,\note{\href{https://orcid.org/0000-0003-0498-0029}{ORCID ID: 0000-0003-0498-0029}}}
\author[b,2]{Giandomenico Palumbo,\note{\href{https://orcid.org/0000-0003-1303-1247}{ORCID ID: 0000-0003-1303-1247}}}
\author[c,3]{Simon Pekar,\note{\href{https://orcid.org/0000-0002-0765-8986}{ORCID ID: 0000-0002-0765-8986}}}
\author[d,e,4]{Alfredo Pérez,\note{\href{https://orcid.org/0000-0003-0989-9959}{ORCID ID: 0000-0003-0989-9959}}}
\author[f,5]{and Stefan Prohazka\note{\href{https://orcid.org/0000-0002-3925-3983}{ORCID ID: 0000-0002-3925-3983}}}

 \affiliation[a]{School of Mathematics and Maxwell Institute for Mathematical Sciences,\\
University of Edinburgh, Peter Guthrie Tait road, Edinburgh EH9 3FD, UK}
 \affiliation[b]{School of Theoretical Physics,\\ Dublin Institute for Advanced Studies, 10 Burlington Road, Dublin 4, Ireland}
 \affiliation[c]{Centre de Physique Théorique – CPHT École polytechnique, \\ CNRS Institut Polytechnique de Paris, 91120 Palaiseau Cedex, France}
 \affiliation[d]{Centro de Estudios Científicos (CECs), Avenida Arturo Prat 514, Valdivia, Chile}
 \affiliation[e]{Facultad de Ingeniería, Arquitectura y Diseño, Universidad San Sebastián,\\
sede Valdivia, General Lagos 1163, Valdivia 5110693, Chile}
\affiliation[f]{University of Vienna, Faculty of Physics, Mathematical Physics,\\ Boltzmanngasse 5, 1090, Vienna, Austria}

\emailAdd{j.hartong@ed.ac.uk}
\emailAdd{giandomenico.palumbo@gmail.com}
\emailAdd{simon.pekar@polytechnique.edu}
\emailAdd{alfredo.perez@uss.cl}
\emailAdd{stefan.prohazka@univie.ac.at}

\abstract{We study dipole Chern--Simons theory with and without a cosmological constant in $2+1$ dimensions. We write the theory in a second order formulation and show that this leads to a fracton gauge theory coupled to Aristotelian geometry which can also be coupled to matter. This coupling exhibits the remarkable property of generalizing dipole gauge invariance to curved spacetimes, without placing any limitations on the possible geometries. We also use the second order formulation to construct a higher dimensional generalization of the action. Finally, for the $(2+1)$-dimensional Chern--Simons theory we find solutions and interpret these as electric monopoles, analyze their charges and argue that the asymptotic symmetries are infinite-dimensional.}

\begin{document} 
\maketitle
\flushbottom

\section{Introduction}
\label{sec:introduction}

Fractons~\cite{Chamon:2004lew,Haah:2011drr} are novel, at this point
theoretical, quasiparticles with the distinctive feature of having
only limited
mobility~\cite{Nandkishore:2018sel,Pretko:2020cko,Grosvenor:2021hkn}. Their
underlying (exotic) dipole symmetry falls into the broader class of
generalized symmetries~\cite{Brauner:2022rvf,Cordova:2022ruw} that
challenge, and hence improve, our understanding of quantum field
theories.

One puzzling aspect is their coupling to
spacetime~\cite{Gromov:2017vir,Slagle:2018kqf}. While the matter
fields~\cite{Pretko:2018jbi} can be coupled to generic Aristotelian
geometry~\cite{Bidussi:2021nmp,Jain:2021ibh} the gauge
theory~\cite{Pretko:2016kxt,Pretko:2016lgv} that mediates the forces
puts restrictions on the admissible
spacetimes~\cite{Slagle:2018kqf,Bidussi:2021nmp,Jain:2021ibh}. The
reason for the restriction lies in the tension between general
Aristotelian covariance and dipole gauge symmetries.  In Cartesian
coordinates the latter gauge transformations with parameter $\Lambda$
act on the gauge fields $\phi$ and symmetric tensor $A_{ij}$ as
\begin{align}
  \label{eq:gauge-trans-orig}
  \vd_{\Lambda} \phi &= \pd_{t}\Lambda &   \vd_{\Lambda} A_{ij}&= -\pd_{i}\pd_{j}\Lambda \,,
\end{align}
where $i,j$ are spatial indices. They couple to matter via
$\rho\delta\phi + J^{ij}\delta A_{ij}$, which leads to the
conservation equation
\begin{align}
  \label{eq:conservation}
  \pd_{t}\rho + \pd_{i}\pd_{j}J^{ij} =0 \, ,
\end{align}
which is at the heart of many of the interesting properties of this
theory.  It implies for example the conservation of the electric
charge $Q=\int \rho dx$ and the dipole moment
$\vec D=\int \vec x \rho dx$ and that isolated monopoles cannot
move. These relations also show that this theory is nonlorentzian and
that generalizing it to generic curved spacetimes is nontrivial.

In this work we will show that it is possible to couple fracton gauge
fields consistently to a particular Aristotelian theory of
gravity.\footnote{For complementary approaches, see, e.g.,
  \cite{Pena-Benitez:2021ipo,Bertolini:2023juh,Bertolini:2023sqa,Pena-Benitez:2023aat,Afxonidis:2023pdq,Jain:2024ngx}.}
We circumvent the earlier no-go results by providing another gauge
theory, which derives from gauging the fracton/dipole algebra in a
spirit similar to the gauging of spacetime symmetry algebras to obtain
Einstein gravity in the first order formulation (see,
e.g.,~\cite{Hartong:2015xda,Figueroa-OFarrill:2022mcy}).  In $2+1$
dimensions this leads to a fracton/dipole Chern--Simons (CS)
theory~\cite{Huang:2023zhp}.\footnote{The existence of this CS theory
  already follows from the correspondence of the fracton and Carroll
  symmetries~\cite{Bidussi:2021nmp,Marsot:2022imf,Figueroa-OFarrill:2023vbj}
  and the fact that theories with Carroll symmetry allow for a CS formulation
  \cite{Bergshoeff:2016soe,Matulich:2019cdo}.}  One of our main
results is to translate this theory into second order formulation (see
Section~\ref{sec:second-order} for the definition of all expressions)
\begin{equation}\label{eq:secfin}
\begin{split}
  S[\phi, A_{\mu\nu}, \tau_\mu, h_{\mu\nu}]=\int d^{3}x e&\left(-\mu\phi h^{\mu\nu}R_{\mu\nu}
    + 2\mu K_{\mu\rho}A_{\nu\sigma}\left(h^{\mu\nu}h^{\rho\sigma}-h^{\mu\rho}h^{\nu\sigma}\right) \right. \\
    &\qquad\left.+\frac{\mu_H}{2}\varepsilon^{\rho\sigma\kappa}\tau_\kappa\left(\partial_\rho\tau_\sigma-\partial_\sigma\tau_\rho\right)\right) \,.
\end{split}
\end{equation}
This action provides a coupling of fracton gauge fields
$(\phi,A_{\mu\nu})$ to the Aristotelian geometry given by
$(\tau_{\mu},h_{\mu\nu})$ and can be generalized to generic spacetime
dimension. Since the gauge fields act as Lagrange multipliers for the
geometry, it exhibits similarities to JT
gravity~\cite{Teitelboim:1983ux,Jackiw:1984je} and BF models.

This coupling possesses the remarkable property that the action
remains invariant under the following generalization of dipole gauge
transformations to curved spacetimes, without requiring additional
restrictions on the geometry
\begin{align}
    \delta\phi&=n^\mu\partial_\mu\bar\Lambda & \delta A_{\mu\nu}&=-P_{(\mu}^\rho P_{\nu)}^\sigma\nabla_\rho\partial_\sigma\bar\Lambda \, , 
\end{align}
where
$P_\mu^\rho = h_{\mu\nu} h^{\nu\rho} = \delta_\mu^\rho - n^\rho
\tau_\mu$ is the spatial projector and $n^\mu$ is the vector dual to
the clock form $\tau_\mu$, i.e., $n^\mu \tau_\mu = 1$ and
$n^\mu h_{\mu\nu} = 0$. This implies a generalization of dipole
conservation, i.e., $\partial_\mu\left(eJ^\mu\right)=0$ where $e$ is
the integration measure (analog of $\sqrt{-g}$ in a relativistic
setup) and where $J^\mu$ is the current
\begin{align}
J^\mu=\rho n^{\mu} +\nabla_\nu\left(P_{\rho}^\mu P_{\sigma}^\nu J^{\rho\sigma}\right) \, ,
\end{align}
where for simplicity we have assumed that the Aristotelian
metric-compatible affine connection $\nabla_\mu$ has no torsion. These
are the curved generalizations of \eqref{eq:gauge-trans-orig} and
\eqref{eq:conservation} (without linearization or further restrictions
on the geometry). Hence dipole gauge invariance puts no restrictions
on the geometry (in any dimension).

Following~\cite{Bidussi:2021nmp,Jain:2021ibh} we show how we can
couple the $(2+1)$-dimensional theory to matter
theories~\cite{Pretko:2018jbi} (Section~\ref{sec:coupling-matter}) and
how we can add a cosmological constant term.  In $2+1$ dimensions we
also construct a solution to the nonlinear equations and derive for
negative and vanishing cosmological constant the conserved
charges. The geometry is spherically symmetric and has nonzero
electric charge and energy and we therefore interpret it as a
monopole. For negative $\cc$ we also discuss the asymptotic symmetries
which are given by an infinite dimensional enhancement of the fracton
algebra (cf., \cite{Perez:2022kax,Perez:2023uwt}).

This work is structured as follows.  In
Section~\ref{sec:fracton-chern-simons} we introduce the fracton
Chern--Simons theory, i.e., the first order formulation, with and
without cosmological constant term.  In Section~\ref{sec:second-order}
we discuss the underlying Aristotelian geometry and translate to
second order formulation, which we use to generalize the action to
generic dimensions (Section~\ref{sec:higher-dimensions}). In $2+1$
dimensions we show that we can couple the theory to matter fields
(Section~\ref{sec:coupling-matter}).  In
Section~\ref{sec:solutions-charges} we find static circularly
symmetric solutions, interpret them as monopoles and discuss their
charges and asymptotic symmetries.  We close by mentioning various
interesting generalizations (Section~\ref{sec:discussion}).  We have
delegated technical aspects concerning the Aristotelian connection to
Appendix~\ref{app:connection}, its curvature to
Appendix~\ref{app:curvature} and Lie algebraic considerations to
Appendix~\ref{app:algebr-cons}.

\section{Fracton Chern--Simons theory in $2+1$ spacetime dimensions}
\label{sec:fracton-chern-simons}

In this section we introduce a Chern--Simons theory based on the
fracton algebra, with and without cosmological constant.

\subsection{Fracton algebra and its invariant metric}
\label{sec:fracton-algebra-its}

The fracton/dipole algebra~\cite{Gromov:2018nbv} in $2+1$ dimensions
is spanned by the set of generators
$\langle J, H, P_{a}, Q,D_{a} \rangle$, which are the usual generators
of symmetry of Aristotelian spacetime, i.e. spatial rotations, time
and space translations, dual to the angular momentum, energy and
linear momentum, as well as two generators of internal symmetry dual
to electric and dipole charge, respectively. The non-vanishing
commutation relations are given by
\begin{align}
  \label{eq:frac-sym}
  [J,P_{a}] &= \epsilon_{ab} P_{b} &  [J,D_{a}] &= \epsilon_{ab} D_{b} & [P_{a},D_{b}] = \delta_{ab} Q \,,
\end{align}
where $a,b = 1,2$ and $\epsilon_{12} =1$. It is a nonsemisimple
algebra with a nontrivial central extension ($Q$) and a trivial one ($H$). In $2+1$ dimensions there exist
other nontrivial central extensions, but since they do not persist for
generic dimensions we will not consider them.

If we want to use the symmetries~\eqref{eq:frac-sym} to construct a
Chern--Simons theory one usually requires the existence of an
invariant metric, that is a symmetric, $\mathrm{ad}$-invariant,
non-degenerate bilinear form on the Lie algebra. The fact that this
algebra is nonsemisimple makes the existence of such an invariant
metric nontrivial. In contradistinction, for semisimple Lie algebras
there is of course always the Killing form (by Cartan's
criterion). For the case at hand, the existence follows from the
isomorphism of the Carroll and fracton/dipole
algebras~\cite{Bidussi:2021nmp}, and the fact that the Carroll algebra
has an invariant metric in $2+1$
dimensions~\cite{Bergshoeff:2016soe,Matulich:2019cdo}.

For the fracton algebra~\eqref{eq:frac-sym} the most general
invariant metric is given by
\begin{align}
  \langle J, Q \rangle &= \mu  & \langle P_{a},D_{b}\rangle &=-\mu\,\epsilon_{ab} & \langle H, H \rangle &= \mu_{H}\nonumber
  \\
  \langle J,J \rangle &= \chi_{J} & \langle J,H \rangle &=\chi_{JH} & \label{eq:inv-metric}
\end{align}
which is non-degenerate for $\mu \neq 0 \neq \mu_{H}$. We will see below that since $\mu_H\neq 0$, one can without loss of
generality always set $\chi_{JH}$ equal to zero in the CS action.

\subsection{Fracton/dipole CS action}
\label{sec:fracton-cs-action}

With these ingredients we can write a Chern--Simons theory
\begin{align}
  S_{\mathrm{CS}}[A] &= \int \langle A \wedge \dd A + \tfrac{1}{3} [A,A] \wedge A \rangle \equiv \int L_{\mathrm{CS}} \,,
\end{align}
for the Lie algebra valued one-form $A$, decomposed as
\begin{align}
  \label{eq:Aconnect}
  A = A_{t} dt + A_{i} dx^{i}= \tau H + e^{a} P_{a} + \omega J + a Q + A^{a} D_{a} \,.
\end{align}
The Chern--Simons action for the fracton algebra~\eqref{eq:frac-sym}
with invariant metric~\eqref{eq:inv-metric} is given by
\begin{align}\label{eq:CSaction}
    S[\tau,e,\omega,a,A] & = \int 2 \mu \left( \omega \wedge da - \epsilon_{ab}e^{a} \wedge d A^{b} + e^{a} \wedge A^{a} \wedge \omega \right)+ \mu_{H} \tau \wedge d \tau \nonumber\\
    &\qquad + \chi_{J} \omega\wedge d\omega + 2 \chi_{JH} \omega \wedge d \tau \,.
\end{align}
This theory was already discussed in~\cite{Huang:2023zhp}, where the
term proportional to $\mu_{H}$ was mentioned, but left implicit. Since
$\tau$ is a relevant part of the Aristotelian geometry we will keep it
explicit.

Using the fact that $\mu_H$ was assumed to be nonzero, we can rewrite the last
three terms of the above action as
\begin{equation}
    \mu_{H} \tau \wedge d \tau + \chi_{J} \omega\wedge d\omega + 2 \chi_{JH} \omega \wedge d \tau = \mu_H\left(\tau+\alpha\omega\right)\wedge d\left(\tau+\alpha\omega\right)+\beta\omega\wedge d\omega \,,
\end{equation}
(up to a total derivative) where $\alpha$ and $\beta$ are given by
\begin{align}
    \alpha &= \chi_{JH}/\mu_H &  \beta&=\chi_J-\chi^2_{JH}/\mu_H \,.
\end{align}
By performing a redefinition of $\tau$, given by
$\tau'=\tau+\alpha\omega$, we can remove the term with $\chi_{JH}$
entirely, and so without loss of generality we can set
\begin{equation}
    \chi_{JH}=0 \,.
\end{equation}
For simplicity we will
furthermore assume that
\begin{equation}
    \chi_{J} = 0 \,.
\end{equation}

Like every CS theory in $2+1$ dimensions, this theory has no local propagating degrees of
freedom and,  without further input, it does not depend on any
(non)lorentzian metric or geometry. In the next sections we will
however interpret some of these generators in terms of Aristotelian
geometry~\cite{Bidussi:2021nmp,Jain:2021ibh}, e.g., we will impose
additional restrictions on the vielbeine
(see~\eqref{eq:inversevielbeine}). This means we introduce additional
structure, which does not change the degrees of freedom, but the
theory depends then on geometric quantities, like the Aristotelian
analog of a metric.

\subsection{Equations of motion, and gauge transformations}
\label{sec:equations-motion}

The equations of motion are given by the usual curvature equals zero
equations where the curvature is
\begin{equation}
\label{eq:CSflat}
  F=\dd A + \tfrac{1}{2} [A,A]=0 \, .
\end{equation}
In components, if we vary the fields in the action, this amounts to
\begin{subequations}
  \begin{align}
    \vd \tau&:&  d\tau &= 0 \\
    \vd e^{a}&:&  dA^{a}- \epsilon_{ab}\,\omega \wedge A^{b} &= 0 \\
     \vd a&:&  d\omega &= 0 \\
    \vd \omega&:&  da  + e^{a} \wedge A^{a} &= 0\label{eq:EOMomega} \\
    \vd A^{a}&:&  de^{a} - \epsilon_{ab}\,\omega \wedge e^{b} &= 0 \,.\label{eq:EOMAa}
  \end{align}
\end{subequations}
The gauge transformations are of the form
$\vd A= \dd \varepsilon + [A,\varepsilon]$ where
\begin{equation}
    \varepsilon = \lambda J+\zeta H+\zeta^a P_a+\Lambda Q+\Lambda^a D_a\,.
\end{equation}
In components this reads
\begin{subequations}
  \begin{align}
    \delta\tau & =  d\zeta \label{eq:gaugetrafotau}\\
    \delta e^a & =  d\zeta^a+\lambda\,\epsilon^a{}_b\,e^b-\omega\,\epsilon^a{}_b\,\zeta^b \label{eq:gaugetrafoe}\\
    \delta \omega & =  d\lambda \\
    \delta a & =  d\Lambda+e^a\,\Lambda_a-A^a\,\zeta_a \label{eq:gaugetrafoa}\\
    \delta A^a & =  d\Lambda^a+\lambda\,\epsilon^a{}_b\, A^b-\omega\,\epsilon^a{}_b\,\Lambda^b\,.
  \end{align}
\end{subequations}

In Section~\ref{sec:second-order} we provide a more detailed and
general analysis, but let us first give some intuition on how we can
recover the dipole conservation~\eqref{eq:conservation} from the
coupling to the fields $a$ and $A^{a}$ of the CS theory. We restrict ourselves to a flat background in Cartesian coordinates, i.e.,
$e_\mu{}^a = \delta^{a}_{\mu}$, $\tau_\mu = \delta_\mu^t$ and
$\omega=0$, which allows to simply replace tangent indices $a$, $b$
into spatial ones $i$, $j$. The charge and dipole gauge
transformations are then given by
\begin{align}
    \delta a_{i} & =  \pd_{i}\Lambda+\Lambda_i & \delta a_t & =  \pd_t\Lambda &  \delta A_i{}^j & =  \pd_{i}\Lambda^j & \delta A_t{}^j & =  \pd_t \Lambda^j
\end{align}
and when we set $\Lambda_i=-\pd_{i}\Lambda$ (which is the residual gauge transformation of the gauge choice $a_i=0$) we find that $a_t$ and
$A_{ij}$ transform precisely like the gauge fields
in~\eqref{eq:gauge-trans-orig}, which implies the dipole conservation
law.

\subsection{Adding a cosmological constant}
\label{sec:cosmological-constant}

We can deform the fracton algebra~\eqref{eq:frac-sym} by adding
curvature (``cosmological constant'') $\cc$, which results in the
algebra (see Appendix~\ref{app:algebr-cons} for the details)
\begin{align}
  [J,P_{a}] &= \epsilon_{ab} P_{b} &  [J,D_{a}] &= \epsilon_{ab} D_{b} & [P_{a},D_{b}] &= \delta_{ab} Q \nonumber \\
  [Q,P_{a}] &= \cc D_a & [P_a, P_b] &= \cc \epsilon_{ab} J \,                      , \label{eq:frac-curved-3d}
\end{align}
with the most general invariant metric
  \begin{align}
    \langle J, Q \rangle &= \mu  & \langle P_{a},D_{b}\rangle &=-\mu\,\epsilon_{ab} & \langle H, H \rangle &= \mu_{H} \nonumber
    \\
    \langle J,J \rangle &= \chi_{J} & \langle P_{a},P_{b} \rangle &= \cc\,\chi_{J}\,\delta_{ab} \,,
  \end{align}
which is non-degenerate for $\mu \neq 0 \neq \mu_{H}$.  With the
connection~\eqref{eq:Aconnect} the CS action~\eqref{eq:CSaction} is
then given by
\begin{align}\label{eq:CSaction-cosm}
    S_{\cc}[\tau,e,\omega,a,A] & = \int 2 \mu \left( \omega \wedge da - \epsilon_{ab}e^{a} \wedge d A^{b} + e^{a} \wedge A^{a} \wedge \omega + \frac{\cc}{2} \epsilon_{ab}e^{a} \wedge e^{b} \wedge a \right) \nonumber\\
                               &\qquad
                                 + \mu_{H} \tau \wedge d \tau
                                 + \chi_{J} 
      \left(
      \omega\wedge d\omega + \cc (e^{a} \wedge de^{a} + \epsilon_{ab}e^{a} \wedge e^{b} \wedge \omega)
      \right)  \,, 
\end{align}
with equations of motions (for $\chi_J = 0$)
\begin{subequations}
  \label{eq:EOM-curved}
  \begin{align}
    \vd \tau&:&  d\tau &= 0 \label{eq:EOML1} \\
    \vd e^{a}&:&  dA^{a}- \epsilon_{ab}\omega \wedge A^{b}+\cc a\wedge e^{a} &= 0  \label{eq:EOML2} \\
     \vd a&:&     d\omega +\frac{1}{2} \cc \epsilon_{ab}e^{a}\wedge e^{b} &= 0 \label{eq:EOML3}\\
    \vd \omega&:&  da  + e^{a} \wedge A^{a} &= 0 \label{eq:EOML4} \\
    \vd A^{a}&:&  de^{a} - \epsilon_{ab}\omega \wedge e^{b} &= 0 \,.\label{eq:EOML5}
  \end{align}
\end{subequations}
Equation \eqref{eq:EOML3} shows that $\cc$ can be interpreted as
adding a cosmological constant to the geometry. 

\section{Second-order formulation}
\label{sec:second-order}

In this section we will translate the Chern--Simons action to the
second-order formulation.  Roughly speaking and similar to general
relativity, we integrate out $\omega$ and find a connection that is
built out of the Aristotelian metric like fields $\tau_{\mu}$ and
$h_{\mu\nu}$. We show that the resulting action is gauge invariant and derive
how to couple it to matter. With the exception of the matter coupling we
show how we can generalize to generic dimension.

\subsection{Integrating out fields}
\label{sec:integr-out-fields}

We will assume that $(\tau_\mu\,,e_\mu{}^a)$ forms an invertible set of
vielbeine whose inverse is given by $(n^\mu\,,e^\mu{}_a)$ where
\begin{equation}\label{eq:inversevielbeine}
    n^\mu \tau_\mu = 1\,,\qquad e_\mu^a e^\nu_a + \tau_\mu n^\nu = \delta_\mu^\nu\,.
\end{equation}
We define $\xi^\mu$ as $\zeta=\xi^\mu\tau_\mu$ and
$\zeta^a=\xi^\mu e_\mu^a$, so that we have a bijective correspondence
between $\xi^\mu$ and $(\zeta,\zeta^a)$. Using the equations of
motion, i.e., that $F=0$ it then follows that the $(\zeta,\zeta^a)$
transformations are on shell equivalent to Lie derivatives along
$\xi^\mu$.

Since $(\tau_\mu\,, e_\mu{}^a)$ are invertible, the
gauge transformation can also be written as
\begin{equation}\label{eq:deltabarA}
    \delta A_\mu=\partial_\mu\varepsilon+\left[A_\mu\,,\varepsilon\right]=\mathcal{L}_\xi A_\mu+\partial_\mu \Sigma+\left[A_\mu\,,\Sigma\right]+\xi^\nu F_{\mu\nu}=\bar\delta A_\mu+\xi^\nu F_{\mu\nu}\,,
\end{equation}
where the last equality defines $\bar\delta A_\mu$ and where
\begin{align}
  \label{eq:Sigma}
    \varepsilon&=\xi^\mu A_\mu+\Sigma & \Sigma&=\bar\lambda J+\bar\Lambda Q+\bar\Lambda^a D_a\,.
\end{align}
Because the difference between $\delta A_\mu$ and $\bar\delta A_\mu$
is proportional to the equations of motion it follows that
$\bar\delta A_\mu$ is also a gauge symmetry of the theory. Basically this
is because we have
$\xi^\sigma\langle F_{[\mu\nu}F_{\rho]\sigma}\rangle=0$. The variation
of the CS action is (up to boundary terms)
\begin{equation}
    \delta S_{\text{CS}}=2\int\langle F\wedge \delta A\rangle\,,
\end{equation}
and the above conclusion follows from
$\langle F\wedge i_\xi F\rangle=0$ where $i_\xi$ denotes the interior
product with respect to the vector $\xi$.

The equations of motion allow us to solve for some of the fields
algebraically in terms of the other fields. If the set of fields we
vary can be solved for that same set of fields algebraically we are
allowed to substitute these back into the action and obtain an
equivalent description in terms of fewer fields. Since the $\omega$
connection is one of these fields the resulting action will be a
second order formulation of the theory.

Consider equations \eqref{eq:EOMomega} and \eqref{eq:EOMAa}. The
latter can be written as
\begin{equation}\label{eq:curvatureP}
    \partial_\mu e_\nu{}^a-\partial_\nu e_\mu{}^a-\epsilon_{ab}\left(\omega_\mu e_\nu{}^b-\omega_\nu e_\mu{}^b\right)=0\,.
\end{equation}
By contracting this equation with $n^\mu$ and $e^\nu{}_c$ we can solve
for $\omega_\mu$ leading to
\begin{subequations}
  \begin{align}
    n^\mu\omega_\mu  & = \frac{1}{2}\epsilon_{ac}n^\mu e^\nu{}_c\left(\partial_\mu e_\nu{}^a-\partial_\nu e_\mu{}^a\right)\label{eq:omega1}\\
    e^\mu{}_a\omega_\mu & = \frac{1}{2}\epsilon_{cd}e^\mu_c e^\nu{}_d\left(\partial_\mu e_\nu{}^a-\partial_\nu e_\mu{}^a\right)\,.\label{eq:omega2}
  \end{align}
\end{subequations}
Equation \eqref{eq:EOMomega} can be written as
\begin{equation}\label{eq:curvatureQ}
    \partial_\mu a_\nu - \partial_\nu a_\mu + e_\mu{}^a A_\nu{}^a - e_\nu{}^a A_\mu{}^a=0\,,
\end{equation}
from which it follows that
\begin{subequations}
  \begin{align}
    n^\mu A_\mu{}^a & =  n^\mu e^\nu{}_a\left(\partial_\mu a_\nu-\partial_\nu a_\mu\right) \\
    e^\mu{}_b A_\mu{}^a - e^\mu{}_a A_\mu{}^b & =  -e^\mu{}_a e^\nu{}_b\left(\partial_\mu a_\nu-\partial_\nu a_\mu\right)\,.
  \end{align}
\end{subequations}
The second equation tells us that the most general solution to
$A_\mu^a$ is given by
\begin{equation}
\label{eq:Sdef}
    A_\mu{}^a = \frac{1}{2} e^\nu{}_a (\delta_\mu^\rho + n^\rho \tau_\mu)\left(\partial_\rho a_\nu-\partial_\nu a_\rho\right) + S^{a}{}_b e_\mu{}^b =: \tilde A_\mu{}^a + S^{a}{}_b e_\mu{}^b \,,
\end{equation}
where $S^{ab}$ is symmetric in $a$ and $b$, but otherwise arbitrary,
and the $a,b,\ldots$ indices are raised and lowered with a Kronecker
delta, and where we defined $\tilde A_\mu{}^a$. The solution for
$\omega$ is equivalent to imposing \eqref{eq:EOMAa} whereas the
solution for $A^a=\tilde A^a+S^a{}_b e^b$ where $\tilde A^a$ obeys
\eqref{eq:EOMomega}. Using the solutions for $\omega$ and $A^a$ the
Lagrangian can be rewritten (up to a total derivative) as
\begin{eqnarray}
    L_{\mathrm{CS}} & = & 2\mu\left(\omega\wedge\left(da+e^a\wedge A^a\right)-\epsilon_{ab}de^a\wedge A^b\right)+\mu_H\tau\wedge d\tau\nonumber\\
    & = & 2\mu\left(-\epsilon_{ab}de^a\wedge S^b{}_c e^c-\epsilon_{ab}de^a\wedge\tilde A^b\right)+\mu_H\tau\wedge d\tau\nonumber\\
    & = & 2\mu\left(-\epsilon_{ab}de^a\wedge S^b{}_c e^c- \omega\wedge e^b\wedge\tilde A^b\right)+\mu_H\tau\wedge d\tau\nonumber\\
    & = & 2\mu\left(-\epsilon_{ab}de^a\wedge S^b{}_c e^c+ \omega\wedge da\right)+\mu_H\tau\wedge d\tau\nonumber\\
    & = & 2\mu\left(a\wedge d\omega-\epsilon_{ab}de^a\wedge S^b{}_c e^c\right)+\mu_H\tau\wedge d\tau\,,\label{eq:2ndorderaction_v1}
\end{eqnarray}
where $\omega$ is no longer an independent connection, but where $S^{ab}=S^{(ab)}$ is an independent variable (as are $\tau_\mu, e_\mu{}^a$ and $a_\mu$).

\subsection{Aristotelian geometry}
\label{sec:arist-geom}

The goal is to rewrite \eqref{eq:2ndorderaction_v1} in terms of an
affine connection and its associated curvature as well as
possibly torsion terms of said affine connection. In order to
introduce such a connection we invoke the following vielbein postulate
\begin{subequations}
  \begin{align}
    0 & =  \partial_\mu\tau_\nu-\Gamma^\rho_{\mu\nu}\tau_\rho\\
    0 & =  \partial_\mu e_\nu{}^a - \epsilon^{ab}\omega_\mu e_{\nu}{}_b -\Gamma^\rho_{\mu\nu}e_\rho{}^a \,.
  \end{align}
\end{subequations}
If we solve these two equations for $\Gamma^\rho_{\mu\nu}$ we obtain
\begin{equation}\label{eq:Gamma_1}
\Gamma^\rho_{\mu\nu}=n^\rho\partial_\mu\tau_\nu+e^\rho{}_b\left(\partial_\mu e_\nu{}^b-\epsilon^{bc}\omega_\mu e_{\nu}{}_c\right)\,.
\end{equation}
It can be shown (see appendix \ref{app:connection}) that for
$\omega_\mu$ given in \eqref{eq:omega1} and \eqref{eq:omega2} we can
write the affine connection as
\begin{equation}\label{eq:Gamma_2}
    \Gamma^\rho_{\mu\nu}=n^\rho\partial_\mu \tau_\nu+\frac{1}{2}h^{\rho\sigma}\left(\partial_\mu h_{\nu\sigma}+\partial_\nu h_{\mu\sigma}-\partial_\sigma h_{\mu\nu}\right)-h^{\rho\sigma}\tau_\nu K_{\mu\sigma}\,,
\end{equation}
where we defined
\begin{align}
    h_{\mu\nu}&=\delta_{ab}e_\mu{}^a e_\nu{}^b &  h^{\mu\nu}&=\delta^{ab}e^\mu{}_a e^\nu{}_b\,,
\end{align}
as well as
\begin{equation}
    K_{\mu\nu}=\frac{1}{2}\mathcal{L}_n h_{\mu\nu}\,.
\end{equation}
This connection is metric compatible in the sense that
\begin{align}
    \nabla_\mu\tau_\nu&=0 & \nabla_\mu h_{\nu\rho}&=0\,,
\end{align}
which follows from the vielbein postulates and is thus true by
design. This also implies that
$\nabla_\mu n^\nu = \nabla_\mu h^{\nu\rho} = 0$. Furthermore, it has
nonzero torsion. Explicitly the torsion is given by
\begin{equation}\label{eq:torsion}
T^\rho_{\mu\nu}=2\Gamma^\rho_{[\mu\nu]}=n^\rho\left(\partial_\mu\tau_\nu-\partial_\nu\tau_\mu\right)+h^{\rho\sigma}\left(\tau_\mu K_{\nu\sigma}-\tau_\nu K_{\mu\sigma}\right)\,.
\end{equation}
The torsion is thus determined by $d\tau$ and $K_{\mu\nu}$. One could
formulate this as saying that the torsion is equal to the intrinsic
torsion of an Aristotelian geometry
\cite{Figueroa-OFarrill:2020gpr}. Intrinsic torsion loosely speaking
is a torsion tensor that is constructed from the geometric data
$\tau_\mu$ and $h_{\mu\nu}$ that is first order in derivatives.

The Riemann tensor associated with this affine connection is
\begin{equation}
    R_{\mu\nu\sigma}{}^\rho=-\partial_\mu\Gamma^\rho_{\nu\sigma}-\Gamma^\rho_{\mu\lambda}\Gamma^\lambda_{\nu\sigma}-(\mu\leftrightarrow\nu)\,.
\end{equation}
A straightforward calculation tells us that
\begin{equation}\label{eq:2DRiemann}
    R_{\mu\nu\sigma}{}^\rho=\epsilon^{bc}e^\rho{}_b e_{\sigma}{}_c \left(\partial_\mu\omega_\nu-\partial_\nu\omega_\mu\right)\,.
\end{equation}
The Ricci tensor is defined as $R_{\mu\sigma}=R_{\mu\rho\sigma}{}^\rho$. It follows that 
\begin{equation}
    h^{\mu\sigma}R_{\mu\sigma}=\epsilon^{bc}e^\rho{}_b e^\mu{}_c\left(\partial_\mu\omega_\rho-\partial_\rho\omega_\mu\right)\,.
\end{equation}

\subsection{Fracton gauge fields on an Aristotelian geometry}
\label{sec:fracton-gauge-fields}

From the torsion constraint
$de^a - \epsilon^{ab} \omega \wedge e_b = 0$ we can deduce (by
applying the exterior differential) that
\begin{align}
    d^2 e^a - \epsilon^{ab} d \omega \wedge e_b + \epsilon^{ab} \omega \wedge d e_b &= 0 &  &\Leftrightarrow  & e^a \wedge d \omega &= 0 \,.
\end{align}
Using this it can be shown that the first term on the last line of
\eqref{eq:2ndorderaction_v1} can be written as
\begin{equation}
    a\wedge d\omega=\phi\tau\wedge d\omega=\frac{\phi}{2}e^\rho{}_a e^\sigma{}_b\left(\partial_\rho\omega_\sigma-\partial_\sigma\omega_\rho\right)\tau\wedge e^a\wedge e^b=-\frac{\phi}{2}h^{\mu\nu}R_{\mu\nu}\tau\wedge e^1\wedge e^2\,,
\end{equation}
where $\phi$ is given by $\phi=n^\mu a_\mu$. In other words
we decompose the gauge potential $a_\mu$ as
\begin{equation} \label{eq:a-decomposition}
    a_\mu = \phi \tau_\mu + \phi_a e_\mu{}^a \,,
\end{equation}
where $\phi = n^\mu a_\mu$ and $\phi_a = e^\mu{}_a a_\mu$. In order to
rewrite the second term on the last line of
\eqref{eq:2ndorderaction_v1} we use that
\begin{equation}
    K_{ab}=e^\mu{}_a e^\nu{}_b K_{\mu\nu}=\frac{1}{2}n^\rho e^\sigma{}_a\left(\partial_\rho e_{\sigma}{}_b - \partial_\sigma e_{\rho}{}_b \right)+ (a\leftrightarrow b)\,.
\end{equation}
Using this we can write
\begin{equation}
    de^a\wedge e^c\epsilon_{ab}\,S^b{}_c = K^{ad}S^{bc}\left(\delta_{ad}\delta_{bc}-\delta_{bd}\delta_{ac}\right)\tau\wedge e^1\wedge e^2\,.
\end{equation}
The last term on the last line of \eqref{eq:2ndorderaction_v1} can be written as
\begin{equation}
    \tau\wedge d\tau=\frac{1}{2}\epsilon^{ab}e^\rho{}_a e^\sigma{}_b\left(\partial_\rho\tau_\sigma-\partial_\sigma\tau_\rho\right)\tau\wedge e^1\wedge e^2=\frac{1}{2}\varepsilon^{\rho\sigma\kappa}\tau_\kappa\left(\partial_\rho\tau_\sigma-\partial_\sigma\tau_\rho\right)\tau\wedge e^1\wedge e^2\,,
\end{equation}
where
$\varepsilon^{\rho\sigma\kappa}=e^{-1}\epsilon^{\rho\sigma\kappa}$
with $\epsilon^{\rho\sigma\kappa}$ the Levi-Civita symbol and
$e=\text{det}(\tau_\mu,e_\mu{}^a)$. Hence, we obtain the following
expression for the Lagrangian \eqref{eq:2ndorderaction_v1}
\begin{equation}
    L_{\mathrm{CS}} =\left(-\mu\phi h^{\mu\nu}R_{\mu\nu}-4\mu K^{ad}S^{bc} \delta_{d[a}\delta_{b]c} + \frac{\mu_H}{2}\varepsilon^{\rho\sigma\kappa}\tau_\kappa\left(\partial_\rho\tau_\sigma-\partial_\sigma\tau_\rho\right)\right)\tau\wedge e^1\wedge e^2\,.
\end{equation}

Let us define the following symmetric tensor
\begin{equation}
  \label{eq:Amunu}
    A_{\mu\nu} = e_\mu{}^a e_\nu{}^b S_{ab} = A_\rho{}^a P_{(\mu}^\rho e_{\nu)a} \,,
\end{equation}
where
$P_\mu^\rho = h_{\mu\nu} h^{\nu\rho} = \delta_\mu^\rho - \tau_\mu
n^\rho$. We can then finally write the action as
\begin{equation}\label{eq:secondorderaction}
\begin{split}
  S[\phi, A_{\mu\nu}, \tau_\mu, h_{\mu\nu}]=\int d^{3}x\ e\,&\Big({-\mu}\phi h^{\mu\nu}R_{\mu\nu}
    + 2\mu K_{\mu\rho}A_{\nu\sigma}\left(h^{\mu\nu}h^{\rho\sigma}-h^{\mu\rho}h^{\nu\sigma}\right) \\
    &\qquad +\frac{\mu_H}{2}\varepsilon^{\rho\sigma\kappa}\tau_\kappa\left(\partial_\rho\tau_\sigma-\partial_\sigma\tau_\rho\right)\Big) \,.
\end{split}
\end{equation}
This is a gauge invariant coupling of an
Aristotelian geometry to fracton gauge fields.  We will refer to this
action as the second order formulation of the theory given in
\eqref{eq:CSaction}.

We will next consider the gauge symmetries of this theory. In
particular, with a generalization to higher dimensions in mind, we
will try to understand it independently of its Chern--Simons
formulation.  We will use the $\bar\delta A_\mu$
transformations of \eqref{eq:deltabarA} which we repeat here are
defined as
\begin{equation}
    \bar\delta A_\mu=\mathcal{L}_\xi A_\mu+\partial_\mu\Sigma+\left[A_\mu\,,\Sigma\right]\,,
\end{equation}
where $\Sigma$ is given by \eqref{eq:Sigma}.
In components we have
\begin{equation}
    \bar\delta a_\mu=\mathcal{L}_\xi a_\mu+\partial_\mu\bar\Lambda+e_\mu{}^a\bar\Lambda^a\,.
\end{equation}
Likewise, we have
\begin{equation}
    \bar\delta e_\mu{}^a=\mathcal{L}_\xi e_\mu{}^a+\bar\lambda\epsilon^a{}_b e_\mu{}^b \,,
\end{equation}
as well as
\begin{equation}
    \bar\delta A_\mu{}^a = \mathcal{L}_\xi A_\mu{}^a + \partial_\mu\bar\Lambda^a+\bar\lambda\epsilon^a{}_b A_\mu{}^b - \omega_\mu\epsilon^a{}_b\bar\Lambda^b\,.
\end{equation}
For the clock form $\tau$ we can write 
\begin{equation}
    \bar\delta\tau_\mu=\mathcal{L}_\xi\tau_\mu\,.
\end{equation}
Using the definition of the inverse vielbeine
\eqref{eq:inversevielbeine} we find
\begin{subequations}
  \begin{align}
    \bar\delta e^\mu{}_a & =  \mathcal{L}_\xi e^\mu{}_a+\bar\lambda\epsilon_a{}^b e^\mu{}_b\\
    \bar\delta n^\mu & =  \mathcal{L}_\xi n^\mu\,.
  \end{align}
\end{subequations}

The field $\phi_a$ does not enter the action. In fact we can gauge-fix
it to be zero. This is because we have
\begin{equation}
    \bar\delta \phi_a = \delta (e^\mu{}_a a_\mu) = e^\mu{}_a \partial_\mu \bar\Lambda + \bar\Lambda_a + \bar\lambda \epsilon_a{}^b \phi_b + \xi^\mu\partial_\mu \phi_a\,,
\end{equation}
so that for $\bar\Lambda_a = - e^\mu{}_a \partial_\mu \bar\Lambda$, we
can set $\phi_a = 0=\bar\delta\phi_a$.

The diffeomorphisms and gauge transformation for the remaining
second-order fields entering the action are
\begin{equation}\label{eq:varphi}
    \bar\delta \phi = \bar\delta\left(n^\mu a_\mu\right) = \mathcal{L}_\xi \phi+n^\mu \partial_\mu \bar\Lambda \,,
\end{equation}
and
\begin{equation}\label{eq:varA}
    \bar\delta A_{\mu\nu} = - P_{(\mu}^\rho P_{\nu)}^\sigma {\nabla}_\rho \partial_\sigma \bar\Lambda \,,
\end{equation}
where we used $\bar\Lambda_a = - e^\mu{}_a \partial_\mu \bar\Lambda$.

Finally we will verify that the second order action is gauge invariant
with respect to the $\bar\Lambda$ gauge transformation.  If we take
the second order action \eqref{eq:secondorderaction} and vary it with
respect to $\bar\Lambda$, i.e., using \eqref{eq:varphi} and
\eqref{eq:varA}, then after performing a few partial
integrations\footnote{Since the connection has torsion it is useful to
  note the following when performing partial integrations
\begin{equation}
    \nabla_\mu X^\mu=e^{-1}\partial_\mu\left(e X^\mu\right)+T^\mu_{\mu\nu}X^\nu\,,
\end{equation}
for any vector $X^\mu$.} and using the identity 
\eqref{eq:identityforgaugeinvaction}, we end up with
\begin{equation}\label{eq:variationbarLambda}
    \delta_{\bar\Lambda} S=-2\mu\int d^3 x e\ \bar\Lambda\, h^{\kappa\rho}h^{\lambda\sigma}K_{\kappa\lambda}\left(R_{\rho\sigma}-\frac{1}{2}h_{\rho\sigma}h^{\alpha\beta}R_{\alpha\beta}\right)\,.
\end{equation}
We can see that this identically zero since 
\begin{equation}\label{eq:identity}
    h^{\kappa\rho}h^{\lambda\sigma}\left(R_{\rho\sigma}-\frac{1}{2}h_{\rho\sigma}h^{\alpha\beta}R_{\alpha\beta}\right)=0\,,
\end{equation}
following from \eqref{eq:2DRiemann}.

If we add the cosmological constant term of section
\ref{sec:cosmological-constant} and go to the second order formulation
we end up with
\begin{equation}
\begin{split}
  S[\phi, A_{\mu\nu}, \tau_\mu, h_{\mu\nu}]=\int d^{3}x\ e\,&\Big({-\mu}\phi (h^{\mu\nu}R_{\mu\nu} - 2 \cc)
    + 2\mu K_{\mu\rho}A_{\nu\sigma}\left(h^{\mu\nu}h^{\rho\sigma}-h^{\mu\rho}h^{\nu\sigma}\right) \\
    &\qquad + \frac{\mu_H}{2}\varepsilon^{\rho\sigma\kappa}\tau_\kappa\left(\partial_\rho\tau_\sigma-\partial_\sigma\tau_\rho\right)\Big) \,.
\end{split}
\end{equation}

\subsection{Generalization to higher dimensions}
\label{sec:higher-dimensions}

It is only in the last step, equation \eqref{eq:identity}, that we
explicitly use that we are in $2+1$ dimensions.  One of the benefits
of the second order formulation \eqref{eq:secondorderaction} is that
it can be straightforwardly generalized to higher dimensions.

Explicitly, if we take the action 
\begin{equation}\label{eq:secondorderaction2}
    S[\phi, A_{\mu\nu}, \tau_\mu, h_{\mu\nu}]=\int d^{d+1}x\ e\,\Big(-\mu\phi h^{\mu\nu}R_{\mu\nu}+2\mu K_{\mu\rho}A_{\nu\sigma}\left(h^{\mu\nu}h^{\rho\sigma} - h^{\mu\rho}h^{\nu\sigma}\right)\Big)+S_{\tau, h}\,,
\end{equation}
where all fields are now defined in $d+1$ dimensions, then if we
modify the gauge transformation of $A_{\mu\nu}$ under $\bar\Lambda$ to
\begin{subequations}
  \begin{align}
    \bar\delta \phi  &= n^\mu \partial_\mu \bar\Lambda \\
    \label{eq:varA_genD}
    \bar\delta A_{\mu\nu} &= - P_{(\mu}^\rho P_{\nu)}^\sigma\left[ {\nabla}_\rho \partial_\sigma \bar\Lambda-\bar\Lambda\left(G_{\rho\sigma}-\frac{1}{d-1}h_{\rho\sigma}h^{\kappa\lambda}G_{\kappa\lambda}\right)\right] \,,
  \end{align}
\end{subequations}
in which we defined 
\begin{equation}
    G_{\mu\nu}=R_{\mu\nu}-\frac{1}{2}h_{\mu\nu}h^{\alpha\beta}R_{\alpha\beta}\,,
\end{equation}
whose spatial projection is a $d$-dimensional Einstein tensor, it
follows that \eqref{eq:secondorderaction2} is gauge invariant under
the $\bar\Lambda$ transformation. Note that in
\eqref{eq:secondorderaction2} we left out the term proportional to
$\mu_H$ in \eqref{eq:secondorderaction}. This is because that term
does not generalise so straightforwardly to higher dimensions. We
replaced it with the action $S_{\tau, h}$ which only depends on the
fields $\tau_\mu$ and $h_{\mu\nu}$. For example we can take for the
action $S_{\tau, h}$ the following \cite{Hansen:2020pqs},
\begin{equation}
  \label{eq:Stauh}
    S_{\tau, h}= \frac{1}{4} \int d^{d+1}x \ e\, h^{\mu\rho}h^{\nu\sigma}\left(\partial_\mu\tau_\nu-\partial_\nu\tau_\mu\right)\left(\partial_\rho\tau_\sigma-\partial_\sigma\tau_\rho\right)\,.
\end{equation}
In principle we could take a Ho\v rava--Lifshitz type action for
$S_{\tau, h}$ whose diffeomorphism invariant formulation can be given
in terms of Aristotelian geometry \cite{Hartong:2015zia}.

If we are in $2+1$ dimensions, i.e., we consider
\eqref{eq:secondorderaction2} for $d=2$ with \eqref{eq:Stauh}, then we
know from the rewriting of the first order action that the $\tau_\mu$
equation of motion (upon using all the other equations of motion)
cannot receive any contributions from the terms in
\eqref{eq:secondorderaction2} that are proportional to $\mu$. If vary
\eqref{eq:secondorderaction2} in which we use \eqref{eq:Stauh} with
respect to $\tau_\mu$ we get that $\tau_\mu$ must obey
\cite{Hansen:2020pqs},
\begin{equation}\label{eq:TTNC}
    h^{\mu\rho}h^{\nu\sigma}\left(\partial_\mu\tau_\nu-\partial_\nu\tau_\mu\right)\left(\partial_\rho\tau_\sigma-\partial_\sigma\tau_\rho\right)=0\,,
\end{equation}
which is equivalent (in form notation) to $\tau\wedge d\tau=0$. To get
this result it is sufficient to vary $\tau_\mu$ as
$\delta\tau_\mu=\Omega\tau_\mu$ where $\Omega$ is an arbitrary
function (while keeping $h_{\mu\nu}$ fixed). This says that $\tau$
must be hypersurface orthogonal which is less constraining than what
we had for the 3D CS theory in which we found that $d\tau=0$. The
variation of $S_{\tau, h}$ with respect to $h_{\mu\nu}$ vanishes upon
using the condition \eqref{eq:TTNC}. It would be interesting to work
out the equations of motion of \eqref{eq:secondorderaction2} with
\eqref{eq:Stauh} in general dimensions. 

We can generalize \eqref{eq:secondorderaction2} by adding a
cosmological constant. The action becomes
\begin{equation} \label{eq:secondorderaction3}
    S_\cc = \int d^{d+1}x\ e\,\Big({-\mu}\phi (h^{\mu\nu}R_{\mu\nu} - 2 \cc)+2\mu K_{\mu\rho}A_{\nu\sigma}\left(h^{\mu\nu}h^{\rho\sigma}-h^{\mu\rho}h^{\nu\sigma}\right)\Big)+S_{\tau, h}\,,
\end{equation}
where the only modification is the appearance of a ``cosmological
constant'' term $e\,\phi\,\cc$, with
$\cc = \sigma \frac{d(d-1)}{2\ell^2}$ where $\sigma=-1,1$ and $\ell$
is a length (see Appendix~\ref{app:algebr-cons}). The fracton gauge
transformations are then modified to
\begin{subequations}
  \begin{align}
    \bar\delta \phi  &= n^\mu \partial_\mu \bar\Lambda \label{eq:gaugephiccgend} \\
    \delta A_{\mu\nu} &= - P_{(\mu}^\rho P_{\nu)}^\sigma\left[ \nabla_\rho \partial_\sigma\bar\Lambda - \left(G_{\rho\sigma} - \frac{1}{d-1} h_{\rho\sigma} h^{\alpha\beta} G_{\alpha\beta} - \frac{1}{d-1} h_{\rho\sigma} \cc \right) \bar\Lambda\right] \,.\label{eq:gaugeAccgend}
  \end{align}
\end{subequations}

All theories (in three or higher spacetime dimensions, with and
without cosmological constant) share many similarities with magnetic
Carroll gravity defined in \cite{Bergshoeff:2017btm} and studied,
e.g., in \cite{Campoleoni:2021blr,Hansen:2021fxi}. Besides the issue
of interpreting the different fields entering the action, the main
difference between these two physical situations lies mainly in the
treatment of the clock form and the issue of boost-invariance.  While
magnetic Carroll gravity is a boost-invariant theory for the
Carrollian metric, the equivalent gauge-invariance in fractonic
theories has been exploited to arrive at the transformation laws
\eqref{eq:varA} (or \eqref{eq:varA_genD} or \eqref{eq:gaugeAccgend}).
The clock form in Carroll gravity is a dynamical object while here it
is a fixed part of the geometry, subject to the constraints obtained
by variation of $S_{\tau,h}$.

\subsection{Coupling to matter}
\label{sec:coupling-matter}

In \cite{Bidussi:2021nmp,Jain:2021ibh} it was shown that the complex
scalar field $\Phi$ with global dipole symmetry can be coupled to an
arbitrary curved Aristotelian geometry leading to the following action
(where we adapted the result of \cite{Bidussi:2021nmp} to the notation
used here)
\begin{equation}
\label{eq:curved-scalar-non-inv}
  S_{\text{scalar}}
  =\int d^{d+1}x\ e\left[
  \left( n^\mu \partial_\mu\Phi-i \phi \Phi\right)\left(n^\nu \partial_\nu\Phi^\star + i\phi \Phi^\star\right) -m^2\vert\Phi\vert^2 - \lambda h^{\mu\nu}h^{\rho\sigma}\hat X_{\mu\rho}{\hat X}^\star_{\nu\sigma}
  \right] \, ,
\end{equation}
where 
\begin{equation}
\hat X_{\mu\nu} = P_{(\mu}^\rho P_{\nu)}^\sigma\left(\partial_\rho \Phi \partial_\sigma \Phi - \Phi \nabla_\rho\partial_\sigma\Phi\right)-iA_{\mu\nu}\Phi^2\, ,
\end{equation}
in which $\nabla_\rho$ is covariant with respect to the Aristotelian
connection \eqref{eq:Gamma_2}. The parameters $m^2$ and $\lambda$ are
real numbers.

The Lagrangian \eqref{eq:curved-scalar-non-inv} is gauge invariant
under the gauge transformations
\begin{align}
  \label{eq:curvedgaugetrafo}
    \delta\phi&=n^\mu\partial_\mu\bar\Lambda & \delta A_{\mu\nu}&=-P_{(\mu}^\rho P_{\nu)}^\sigma\nabla_\rho\partial_\sigma\bar\Lambda &  \delta\Phi&=i\bar\Lambda\Phi\,.
\end{align}
Comparing \eqref{eq:gaugephiccgend} and \eqref{eq:gaugeAccgend} with
\eqref{eq:curvedgaugetrafo} we see that the coupling to
\eqref{eq:curved-scalar-non-inv} only works in $2+1$ dimensions (with
$\cc=0$). We can now simply add the actions
\eqref{eq:secondorderaction2} and \eqref{eq:curved-scalar-non-inv} and
set $d=2$ leading to
\begin{eqnarray}
    S & = & \int d^{3}x\ e\,\Big({-\mu}\phi h^{\mu\nu}R_{\mu\nu}+2\mu K_{\mu\rho}A_{\nu\sigma}\left(h^{\mu\nu}h^{\rho\sigma}-h^{\mu\rho}h^{\nu\sigma}\right)+\nonumber\\
    &&+
  \left( n^\mu \partial_\mu\Phi-i \phi \Phi\right)\left(n^\nu \partial_\nu\Phi^\star + i\phi \Phi^\star\right) -m^2\vert\Phi\vert^2 - \lambda h^{\mu\nu}h^{\rho\sigma}\hat X_{\mu\rho}{\hat X}^\star_{\nu\sigma}
  \Big)+S_{\tau, h}\,,
\end{eqnarray}
where $S_{\tau,h}$ is for example given by \eqref{eq:Stauh} or by the
term proportional to $\mu_H$ in \eqref{eq:secondorderaction}.

If we now vary $\phi$ and $A_{\mu\nu}$ we get equations for the
curvature that are determined by the scalar field. Hence, this theory
is not necessarily restricted to maximally symmetric spacetimes (as
was the case in \cite{Slagle:2018kqf,Bidussi:2021nmp} where the dipole
gauge theories were quadratic in the gauge fields).

\section{Solutions and charges}
\label{sec:solutions-charges}

In this Section we will derive circularly symmetric solutions of the
Chern--Simons theory (for any $\cc$). Geometrically they share
similarities with the spatial geometries of $2+1$ dimensional gravity,
but our analysis of the charges shows that they carry electric charge
and we therefore interpret them as monopoles. We also comment on
asymptotic symmetries which infinitely enhance the fracton algebra.

\subsection{Circularly symmetric solutions in $2+1$ dimensions}
\label{sec:circ-symm-solut}

In this section, we will discuss circularly symmetric solutions to the
field equations~\eqref{eq:EOM-curved}, which describe the field
generated by an electric monopole in a curved background with
and without a cosmological constant.

We will make use of the first order formulation. Let us consider a
circularly symmetric ansatz of the form
\begin{align}
  \tau&=N\left(r\right)dt &  e^{1}& =\frac{1}{f\left(r\right)}dr & e^{2}& =rd\theta  & a&=\phi\left(r\right)dt &  S_{ab}&=S\left(r\right)\delta_{ab} \, ,\label{eq:ansatz}
\end{align}
where $\theta$ is $2\pi$ periodic.  We only specify $S_{ab}$, which we
defined as $A^{a} =: \tilde A^{a} + S^{ab} e^{b}$, since $\tilde A$
will be determined by $a$ (as we have already discussed
around~\eqref{eq:Sdef}).

The Aristotelian geometry given by $\tau$ and $e^a$ is completely
determined by \eqref{eq:EOML1}, \eqref{eq:EOML3} and
\eqref{eq:EOML5}. Indeed, the equation of motion \eqref{eq:EOML1}
shows that $N\left(r\right)$ is constant. In particular, by selecting
an appropriate time normalization, $N$ can be set to one, resulting in
\begin{align}
  \tau=dt \, .
\end{align}
Additionally, equations \eqref{eq:EOML3} and \eqref{eq:EOML5} imply
\begin{align}
  \label{eq:omega}
  \omega=f\left(r\right)d\theta \, ,
\end{align}
with
\begin{align}
  f\left(r\right)=\sqrt{-\cc r^{2}-M} \, ,
\end{align}
where $M$ is a real constant.\footnote{Strictly speaking there is the
  freedom to have both signs, i.e., $f(r)=\pm \sqrt{-\cc r^{2}-M}$,
  but since we can absorb this freedom into the orientation of
  $\theta$ in \eqref{eq:omega} we will restrict henceforth to the
  positive root.}

The fractonic fields are determined by \eqref{eq:EOML2} and
\eqref{eq:EOML4}. One finds
\begin{align}
  \label{eq:phiSsol}
  \phi\left(r\right)&=\phi_{0}\sqrt{-\cc r^{2}-M} &  S\left(r\right)&=S_{0} \, ,
\end{align}
where $\phi_{0}$ and $S_{0}$ are integration constants and
$A^{a}=\delta^{a}_{1} \cc r \phi_{0} dt + S_{0}e^{a}$.

In sum, the Aristotelian geometry of the circularly symmetric solution
is described by the following clock form and spatial metric: 
\begin{align}
  \tau&=dt & h_{\mu\nu}dx^{\mu}dx^{\nu}&=\frac{dr^{2}}{-\cc r^{2}-M}+r^{2}d\theta^{2} \, ,
\end{align}
while the fractonic fields are given by (cf.~\eqref{eq:Amunu})
\begin{align}
  \phi=\phi_{0}\sqrt{-\cc r^{2}-M},\qquad\qquad A_{\mu\nu}=S_{0}h_{\mu\nu} \, .
\end{align}
In analogy to their lorentzian geometries we called the integration
constant $M$, but it should not be interpreted as a mass, but rather
as a charge. This can be inferred from the fact that the curvature of
the geometry~\eqref{eq:EOML3} comes from the coupling $a_{\mu}J^{\mu}$
rather than from coupling to $e_{\mu}^{a}$.

Let us first focus on the flat case, which is the well-defined limit
$\cc \to 0$ with metric $-\frac{dr^{2}}{M}+r^{2}d\theta^{2}$ (for the
following remarks further details are, e.g.,
in~\cite{Deser:1983tn,Deser:1988qn} and references therein). For
$M=-1$ this is the plane with flat metric, while for $-1<M<0$ the
plane is deformed into a cone, which is metrically flat except at the
tip which can be interpreted as a point particle. When $M \to 0$ the
geometry approaches a cylinder and when $M<-1$ it is a conical
excesses. When $M>0$ we can think about it as a Milne universe. The
ansatz in \eqref{eq:ansatz} assumes a static and circularly symmetric
configuration, thereby precluding the possibility of deriving a
rotating solution from it. In relativistic gravitational theories, a
common technique to obtain rotating solutions involves applying an
improper boost to a static and circularly symmetric solution. However,
this method is not applicable here due to the absence of
boosts. Nevertheless, one could consider a more general ansatz that is
stationary and invariant under the Killing vector $\partial_{\theta}$,
allowing the metric to include cross terms with $dt d\theta$. We will
explore this possibility in the future.
 
Let us from now on focus on $\cc <0$ where the spatial metric takes
precisely the same form as the spatial metric of a nonrotating BTZ
black hole~\cite{Banados:1992wn,Banados:1992gq} in general relativity
in $2+1$ dimensions. However, the clock form is different since the
clock form does not depend on any integration constant of the spatial
geometry as would have been the case for the lorentzian
geometry. Furthermore, the geometry does not depend on the integration
constants $\phi_{0}$ and $S_{0}$ of the fractonic gauge fields,
meaning there is no backreaction of the gauge fields on the geometry.
This is similar to the case of Einstein gravity in $2+1$ dimensions
coupled to $U(1)$ abelian Chern-Simons fields (see, e.g.,
\cite{Kraus:2006wn}).

The gauge connection associated with this solution is given by
\begin{subequations}
  \begin{align}
    A_{t} & =H+\phi_{0}\sqrt{-\cc r^{2}-M}\,Q+\cc r\phi_{0}\,D_{1} \\
    A_{r} & =\frac{1}{\sqrt{-\cc r^{2}-M}}\;P_{1}+\frac{S_{0}}{\sqrt{-\cc r^{2}-M}}\,D_{1} \\
    A_{\theta} & =r\,P_{2}+\sqrt{-\cc r^{2}-M}\,J+rS_{0}\,D_{2} \, .
  \end{align}
\end{subequations}
In complete analogy with the Chern-Simons formulation of Einstein
gravity, it is possible to gauge away all the dependence on the radial
coordinate $r$, such that the physical information is encoded in an
auxiliary connection
$\mathfrak{a}=\mathfrak{a}_{t}dt+\mathfrak{a}_{\theta}d\theta$, where
$A=h^{-1}\left(d+\mathfrak{a}\right)h$ for some gauge group element
$h$. For the circularly symmetric solution, one explicitly finds that
\begin{align}
  \label{eq:h-neg}
  h=\exp\left[\frac{1}{\sqrt{-\cc}}\text{Arcoth}\left(\sqrt{1+\frac{M}{\cc r^{2}}}\right)\left(P_{1}+S_{0}D_{1}\right)\right] \, ,
\end{align}
where the auxiliary connection is given by
\begin{equation}
  \label{eq:a_aux}
  \mathfrak{a}=\left(H+\sqrt{-M}\phi_{0}Q\right)dt+\sqrt{-M}Jd\theta.
\end{equation}
The removal of the radial dependence via \eqref{eq:h-neg} can only be
achieved for negative values of $\cc$ and $M$. For vanishing
cosmological constant the auxiliary connection takes exactly the same
form as \eqref{eq:a_aux} and the group element $h$ simplifies to
$h=\exp\left[\frac{r}{\sqrt{-M}}\left(P_{1}+S_{0}D_{1}\right)\right]$.

\subsection{Charges and asymptotic symmetries}
\label{sec:charges}

The charges of this theory are related to large gauge transformations.
To determine them we need to find gauge transformations that preserve
the form of the auxiliary connection~\eqref{eq:a_aux}, i.e., we must
find an $\varepsilon$ such that
$d\varepsilon + \left[\mathfrak{a},\varepsilon\right] =0$ and which leads to
non-vanishing charges. The charge associated with these large gauge
transformations can then be obtained using the canonical formalism
\cite{Regge:1974zd} and it is given by the following
expression~\cite{Banados:1994tn}
\begin{align}
  \delta \mathcal{Q}\left[\varepsilon\right]=-2\oint d\theta\left\langle \varepsilon\delta\mathfrak{a}_{\theta}\right\rangle \, .
\end{align}
When a transformation changes the charge it should not be thought of
as a nonphysical gauge redundancy, but as an observable physical
change.

For the case at hand large gauge transformations are generated
by
\begin{align}
  \varepsilon=\bar{\Lambda}Q \, ,
\end{align}
for a constant $\bar{\Lambda}$.\footnote{When $\chi_{J}$ is nonzero
  there are more large general transformations that lead to
  non-vanishing charges, but they will not provide additional
  information since they will be proportional to this charge.} When
$\bar{\Lambda}$ has no functional variation we find the electric
charge
\begin{align}
  \mathcal{Q}\left[\varepsilon\right]=-4\pi\bar{\Lambda}\mu\sqrt{-M} \, .
\end{align}
This shows that the integration constant $M$, which geometrically
shares some similarities with mass (but is not the mass of the
system), is associated to the electric charge of the system.

The total energy can be obtained by considering the charge associated
with time evolution and can be derived from
\begin{align}
  \delta E=2\oint d\theta\left\langle \mathfrak{a}_{t}\delta\mathfrak{a}_{\theta}\right\rangle \,  .
\end{align}
Then, if one assumes that $\delta\phi_{0}=0$, then the energy of
the solution takes the form 
\begin{align}
  E=-2\pi\mu\phi_{0}M  \, .
\end{align}
Therefore the total energy and the electric charge are related
$E \sim \phi_{0}\mathcal{Q}^{2}$. The constant $S_{0}$ does not appear
in the charges. Indeed, since $A_{\mu\nu}=S_{0}h_{\mu\nu}$, the
constant $S_{0}$ can be interpreted as as labeling a particular ground
state of the symmetric tensor $A_{\mu\nu}$.

Note that the components of the auxiliary connection \eqref{eq:a_aux}
are defined along the generators $H$, $J$ and $Q$, which form a set of
commuting generators. This suggests that a natural set of asymptotic
conditions that accommodate this solution could be given by ``soft
hairy asymptotic conditions'', similar to those introduced
in~\cite{Afshar:2016wfy,Grumiller:2016kcp} whose asymptotic symmetry
algebra is given by a set of $U\left(1\right)$ Kac-Moody current
algebras.  This aligns with the fact that the dipole algebra with a
negative cosmological constant is isomorphic, apart from the central
element $H$, to the three-dimensional Poincaré algebra. Indeed, this
isomorphism allows us to map all the known results in
three-dimensional general relativity in flat space to the case of the
dipole algebra with a negative cosmological constant. Indeed, based on
the results in \cite{Coussaert:1995zp}, after a suitable gauge
transformation, it would be possible to write a set of asymptotic
conditions in the flat space analogue of the highest weight gauge
\cite{Barnich:2006av,Afshar:2013vka,Gonzalez:2013oaa}, where
\begin{align}
  a_{\theta}
  =\left(J-\ell P_{1}\right)
  -\frac{\mathcal{M}\left(t,\theta\right)}{2}\left(J+\ell P_{1}\right)-\frac{\mathcal{L}\left(t,\theta\right)}{2}\left(Q+\frac{1}{\ell}D_{2}\right)-\frac{\mathcal{H}\left(t,\theta\right)}{2}H \, .
\end{align}
Here $\ell$ is the $AdS$ radius related to the cosmological constant
by $\cc=-\ell^{-2}$. The asymptotic symmetry algebra is then given
by the three-dimensional BMS algebra with an additional
$U\left(1\right)$ current, which contains the cosmological dipole
algebra as its wedge algebra. The case when the cosmological constant
vanishes is less clear, as the previous asymptotic conditions do not
appear to have a natural flat limit. However, \eqref{eq:a_aux}
suggests that soft hairy asymptotic conditions might naturally be
applicable. We plan to investigate this further in the future.

\section{Discussion}
\label{sec:discussion}

We conclude by recalling that the main result of the paper is a
metric formulation of a fracton gauge theory
\eqref{eq:secondorderaction} obtained from a Chern-Simon action in 2+1 dimensions, with gauge fields $A_{\mu\nu}$ and $\phi$
coupled to dynamical Aristotelian gravitation fields $h_{\mu\nu}$ and
$\tau_\mu$. We want to
emphasize that the invariance of the action under the dipole gauge transformations given by Eqs. \eqref{eq:varphi} and
\eqref{eq:varA} imposes no restrictions on the geometry, in stark contrast to previous no-go results.~\cite{Slagle:2018kqf,Bidussi:2021nmp,Jain:2021ibh}. This
theory was generalized to higher dimensions without
\eqref{eq:secondorderaction2} and with \eqref{eq:secondorderaction3} a
cosmological constant. Additionally, we showed that the three-dimensional theory can be consistently
coupled to fractonic matter fields \eqref{eq:curved-scalar-non-inv}.

This work opens various interesting avenues for further exploration:
\begin{description}
\item[Other multipole symmetries] The tools we have used in this work
  are of course not restricted to dipole symmetries and it could be
  interesting to generalize to higher multipole moments.
\item[Supersymmetrization] An immediate generalization is the
  supersymmetrization of the CS theory. Again using the correspondence
  to Carroll
  symmetries~\cite{Bidussi:2021nmp,Marsot:2022imf,Figueroa-OFarrill:2023vbj}
  it is clear that such a theory exists~\cite{Ravera:2019ize} and it
  could be interesting to generalize the work of
  Huang~\cite{Huang:2023zhp} and ours to this framework.
\item[Fracton BF gravity] There also exist generalizations to
  $(1+1)$-dimensional gravitational
  models~\cite{Grumiller:2020elf,Gomis:2019nih} in particular there is
  an analog proposal for fracton BF gravity \cite{Ecker:2023uwm} to
  which much of what we have done could be applied.
\item[Relation to scalar charge gauge theories] In order to make
  contact with more standard gauge theories of fractons on flat
  space~\cite{Pretko:2016kxt,Pretko:2016lgv} let us take the action
  \eqref{eq:secondorderaction} and choose a background configuration
  which satisfies the equations of motion at 0$^\text{th}$-order,
  given by the flat Aristotelian background $\bar \tau_t = 1$ and
  $\bar h_{ij} = \delta_{ij}$ and all other fields are
  zero. Linearizing the theory \eqref{eq:secondorderaction} up to
  quadratic order around this background yields, among others, the
  term $E^{ij} \dot A_{ij} + \phi\,\partial_i \partial_j E^{ij}$
  (where $E_{ij}$ is related to the metric perturbation
  $e_{ij} = h_{ij} - \bar h_{ij}$ by
  $E_{ij} = e_{ij} - \delta_{ij} e_k{}^k$) which is ubiquitous in the
  Hamiltonian treatment of fracton gauge theories performed, e.g., in
  \cite{Bidussi:2021nmp}. We reserve a more thorough study of the
  Hamiltonian formulation of the theory displayed in
  \eqref{eq:secondorderaction2} or \eqref{eq:secondorderaction3} and
  its relation to the theories (or others) described
  in~\cite{Pretko:2016kxt,Pretko:2016lgv,Bidussi:2021nmp} for future
  works.
\item[Infrared triangle, memory effects] It was recently
  shown~\cite{Perez:2022kax,Perez:2023uwt} that fracton theories allow
  for interesting interrelations, called infrared
  triangle~\cite{Strominger:2017zoo}, between asymptotic symmetries,
  soft theorems and (double kick) memory effects. The fracton CS
  theory also allows for infinite dimensional asymptotic symmetries
  (cf., Section~\ref{sec:charges}) which makes it natural to expect
  related soft theorems and memory effects. What makes the case at
  hand an interesting challenge is that the gauge theory is
  topological and therefore has no propagating degrees of freedom.
\item[Applications to condensed matter systems] We briefly suggest a
  possible application of our fractonic Chern-Simons theory in the
  framework of topological phases of matter. We first observe that our
  action \eqref{eq:CSaction-cosm} can be seen as an one-loop effective
  topological field theory induced by integrating out some massive
  degrees of freedom. In particular, in a $(2+1)$-dimensional
  microscopic system made by non-relativistic massive fermions with
  conserved electric charge, dipole and rotational symmetry but broken
  time-reversal symmetry, \eqref{eq:CSaction-cosm} can describe the
  topological response of the system to external probings. In fact,
  the first term is known in the condensed-matter literature as first
  Wen-Zee term \cite{WZ} and in absence of dipole conservation, it has
  been employed to study several kinds of topological systems, such as
  quantum Hall insulators and higher-order topological phases in two
  space dimensions \cite{Liu,You2,Manjunath2,May-Mann2,Rao}. The
  \emph{first Wen-Zee term}, entirely related to the charge
  conservation and rotational symmetry of the system, gives rise to
  the shift invariant and corresponding Hall viscosity.  On the other
  hand, the first term related to the cosmological constant $\cc$ in
  the same action coincides with the topological response of an atomic
  insulator in two space dimensions \cite{Huang}, which only depends
  on charge conversation and translation symmetry. Finally, the first
  term related to $\chi_J$ in \eqref{eq:CSaction-cosm} is known as
  \emph{second Wen-Zee term} and plays a role mainly in the fractional
  quantum Hall effect
  \cite{Salgado-Rebolledo:2021wtf,Salgado-Rebolledo:2021nfj}.  Thus,
  we expect that our $(2+1)$-dimensional fractonic theory with a
  non-zero cosmological constant represents the low-energy description
  of suitable topological phases augmented by the dipole symmetry,
  namely topological dipole phases (see~\cite{Lam:2024smz} for an example of topological dipole insulator) that will be
  investigated in detail in a future work.
\end{description}

\appendix

\acknowledgments

This work was initiated while StP was supported by the Leverhulme
Trust Research Project Grant (RPG-2019-218) ``What is Non-Relativistic
Quantum Gravity and is it Holographic?''. JH was supported by the Royal
Society University Research Fellowship Renewal ``Non-Lorentzian String
Theory'' (grant number URF\textbackslash R\textbackslash 221038). The
work of SiP was partially supported by the Fonds de la Recherche
Scientifique – FNRS under Grant No. FC.36447. SiP acknowledges the
support of the SofinaBo\"el Fund for Education and Talent and the
\textit{Fonds Friedmann} run by the \textit{Fondation de l'\'Ecole
  polytechnique}. The research of AP is partially supported by Fondecyt grants No 1211226, 1220910 and 1230853.

We acknowledge support from the Erwin Schrödinger International
Institute for Mathematics and Physics (ESI) where some of the research
was undertaken at the ``Carrollian Physics and Holography'' thematic
programme.

\section{Affine connection}
\label{app:connection}

The purpose of this appendix is to show that equation
\eqref{eq:Gamma_2} follows from \eqref{eq:Gamma_1} in which we
substitute \eqref{eq:omega1} and \eqref{eq:omega2}.

First we use completeness to write \eqref{eq:Gamma_1} as
\begin{equation}
\Gamma^\rho_{\mu\nu}=n^\rho\partial_\mu\tau_\nu+e^\rho{}_b\partial_\mu e_\nu{}^b-\epsilon^{bc}e^\rho{}_b e_\nu{}_c\left(\tau_\mu n^\sigma\omega_\sigma+e_\mu{}^d e^\sigma{}_d \omega_\sigma\right)\,.
\end{equation}
In this equation we substitute \eqref{eq:omega1} and \eqref{eq:omega2}
leading to (after some straightforward algebra)
\begin{equation}
\Gamma^\rho_{\mu\nu}=n^\rho\partial_\mu\tau_\nu+h^{\rho\sigma}\tau_\mu K_{\nu\sigma}-h^{\rho\sigma}\tau_\mu h_{\sigma\kappa}\partial_\nu n^\kappa+X^\rho_{\mu\nu}\,,
\end{equation}
where 
\begin{equation}
X^\rho_{\mu\nu}=h^{\rho\lambda}h^{\kappa\sigma}\left(e_{\lambda}{}_b h_{\sigma\mu}\partial_\kappa e_\nu{}^b+e_{\mu}{}_b h_{\sigma\nu}\left(\partial_\kappa e_\lambda{}^b-\partial_\lambda e_\kappa{}^b\right)\right)\,.
\end{equation}
Using completeness once more we can write
\begin{equation}\label{eq:X}
    X^\rho_{\mu\nu}=\tau_\nu n^\alpha X^\rho_{\mu\alpha}+P_\nu^\alpha X^\rho_{\mu\alpha}\,,
\end{equation}
where we have
\begin{equation}
    X^\rho_{\mu\alpha} n^\alpha=-P_\mu^\kappa P_\alpha^\rho\partial_\kappa n^\alpha\,.
\end{equation}
In order to rewrite $P_\nu^\alpha X^\rho_{\mu\alpha}$ we use that 
\begin{equation}
   h^{\rho\lambda}P^\kappa_\nu P^\sigma_\mu\left[ e_{\sigma}{}^b\left(\partial_\kappa e_\lambda{}^b-\partial_\lambda e_\kappa{}^b\right)+\text{cyclic permutations of $\sigma, \kappa, \lambda$}\right]=0\,.
\end{equation}
Applying this identity to one-half of $X^\rho_{\mu\nu}$ while using \eqref{eq:X} for the other half we find
\begin{equation}
    X^\rho_{\mu\nu}=\tau_\nu X^\rho_{\mu\sigma}n^\sigma+\frac{1}{2}h^{\rho\lambda}P^\kappa_\nu P^\sigma_\mu\left(\partial_\kappa h_{\sigma\lambda}+\partial_\sigma h_{\lambda\kappa}-\partial_\lambda h_{\sigma\kappa}\right)\,.
\end{equation}
After a bit of furthermore straightforward algebra we then find
\eqref{eq:Gamma_2}.

\section{Curvature}
\label{app:curvature}

In this appendix we will collect some useful formulas for affine connections $\Gamma^\rho_{\mu\nu}$ with nonzero torsion. 

The covariant derivative will be denoted by $\nabla_\mu$, the Riemann
tensor by ${R}_{\mu\nu\sigma}{}^{\rho}$ and the torsion tensor by
$T^\rho{}_{\mu\nu}$. The latter are defined via
\begin{subequations}
  \begin{align}
    \left[\nabla_\mu,\nabla_\nu\right]X_\sigma & =  R_{\mu\nu\sigma}{}^\rho X_\rho-T^\rho{}_{\mu\nu}\nabla_\rho X_\sigma\\
\left[\nabla_\mu,\nabla_\nu\right]X^\rho & =  -R_{\mu\nu\sigma}{}^\rho X^\sigma-T^\sigma{}_{\mu\nu}\nabla_\sigma X^\rho
  \end{align}
\end{subequations}
from which it follows that
\begin{subequations}
  \begin{align}
{R}_{\mu\nu\sigma}{}^{\rho}&\equiv-\partial_{\mu}{\Gamma}_{\nu\sigma}^{\rho}+\partial_{\nu}{\Gamma}_{\mu\sigma}^{\rho}-{\Gamma}_{\mu\lambda}^{\rho}{\Gamma}_{\nu\sigma}^{\lambda}+{\Gamma}_{\nu\lambda}^{\rho}{\Gamma}_{\mu\sigma}^{\lambda}\label{eq:Riemann_tensor}\\
T^\rho{}_{\mu\nu} &\equiv 2\Gamma^\rho_{[\mu\nu]}\,.\label{eq:torsion_tensor}
  \end{align}
\end{subequations}
The algebraic and differential Bianchi identities are 
\begin{eqnarray}
    R_{[\mu\nu\sigma]}{}^\rho & = & T^\lambda{}_{[\mu\nu}T^\rho{}_{\sigma]\lambda}-\nabla_{[\mu}T^\rho{}_{\nu\sigma]}\\
    \nabla_{[\lambda}R_{\mu\nu]\sigma}{}^\kappa & = & T^\rho{}_{[\lambda\mu}R_{\nu]\rho\sigma}{}^\kappa\,.\label{eq:diffBianchi}
\end{eqnarray}

The Ricci tensor is defined as
\begin{equation}\label{eq:Ricci_tensor_LC}
R_{\mu\nu} \equiv {R}_{\mu\rho\nu}{}^{\rho}\,.
\end{equation}
Our connection satisfies the property that
\begin{equation}
    \Gamma^{\rho}_{\mu\rho}=\partial_\mu\log e\,,
\end{equation}
where $e=\text{det}(\tau_\mu\,, e^a_\mu)$ is the integration measure. From this it follows that
\begin{equation}
R_{\mu\nu\rho}{}^\rho=0\,.
\end{equation}
The antisymmetric part of the Ricci tensor is then
\begin{equation}
    2R_{[\mu\nu]}=T^\lambda{}_{\mu\nu}T^\rho{}_{\lambda\rho}+
    \nabla_\mu T^\rho{}_{\nu\rho}-\nabla_\nu T^\rho{}_{\mu\rho}+\nabla_\rho T^\rho{}_{\mu\nu}\,.
\end{equation}

Consider the differential Bianchi identity \eqref{eq:diffBianchi} and
contract $\kappa$ with $\nu$. Contracting the resulting identity with
$n^\lambda h^{\mu\sigma}$ and using the torsion tensor
\eqref{eq:torsion} leads to
\begin{eqnarray}
    0 & = & \nabla_\lambda\left(n^\lambda h^{\mu\sigma}R_{\mu\sigma}\right)+2h^{\kappa\sigma}h^{\rho\lambda}R_{\rho\sigma}K_{\kappa\lambda}\label{eq:identityforgaugeinvaction}\\
    &&-2\left(h^{\kappa\sigma}h^{\rho\lambda}-h^{\kappa\rho}h^{\lambda\sigma}\right)\left[\nabla_\kappa\nabla_\rho K_{\lambda\sigma}-\nabla_\kappa\left(K_{\lambda\sigma}\mathcal{L}_n\tau_\rho\right)+K_{\lambda\sigma}\mathcal{L}_n\tau_\kappa\mathcal{L}_n\tau_\rho-\nabla_\rho K_{\lambda\sigma}\mathcal{L}_n\tau_\kappa\right]\,.\nonumber
\end{eqnarray}
This identity is used in Section \ref{sec:fracton-gauge-fields} to prove gauge invariance of the second order theory.

\section{Fracton algebraic considerations}
\label{app:algebr-cons}

We start by defining the (anti) de Sitter Carroll algebras in generic
spacetime dimension $d+1$
\begin{equation}
  \begin{aligned}\relax
    [J_{ab}, J_{cd}] &= \delta_{bc} J_{ad} - \delta_{ac} J_{bd} - \delta_{bd} J_{ac} + \delta_{ad} J_{bc}\\
    [J_{ab}, B_c] &= \delta_{bc} B_a - \delta_{ac} B_b\\
    [J_{ab}, P_c] &= \delta_{bc} P_a - \delta_{ac} P_b
  \end{aligned}
  \qquad\qquad
  \begin{aligned}\relax
    [B_a, P_b] &= \delta_{ab} H\\
    [H,P_{a}] &= - \cc B_a \\
    [P_a, P_b] &= - \cc J_{ab} \, .
  \end{aligned}
\end{equation}
where $\cc = \frac{\sigma}{\ell^2}$ with $\sigma = 1$ ($-1$) for
(anti) de Sitter Carroll and is related to the cosmological constant
$\frac{d(d-1)}{2} \cc$ (we follow
\cite{Figueroa-OFarrill:2018ilb,Campoleoni:2022ebj}).

We now replace the Carroll boost by dipole moment
$B_{a} \mapsto D_{a}$ and Carroll energy by charge $H \mapsto -Q$ and
add a central element $H$ to the new algebra. This curved fracton
algebra is then spanned by
$\mathfrak{g} = \langle J_{ab},H,P_{a},Q,D_{a}\rangle$ and given by
\begin{equation}
  \label{eq:frac-curved-d}
  \begin{aligned}\relax
    [J_{ab}, J_{cd}] &= \delta_{bc} J_{ad} - \delta_{ac} J_{bd} - \delta_{bd} J_{ac} + \delta_{ad} J_{bc}\\
    [J_{ab}, D_c] &= \delta_{bc} D_a - \delta_{ac} D_b\\
    [J_{ab}, P_c] &= \delta_{bc} P_a - \delta_{ac} P_b
  \end{aligned}
  \qquad\qquad
  \begin{aligned}\relax
    [P_a, D_b] &= \delta_{ab} Q\\
    [Q,P_{a}] &=  \cc D_a \\
    [P_a, P_b] &= - \cc J_{ab} \, .
  \end{aligned}
\end{equation}
To obtain further geometric understanding let us think about the
homogeneous space where we quotient $\mathfrak{g}$ by
$\mathfrak{h}=\langle J_{ab}, Q, D_{a} \rangle$. The homogeneous space
is a curved Aristotelian homogeneous space\footnote{See,
  e.g.,~\cite{Figueroa-OFarrill:2018ilb} for a precise definition.}
either $\mathbb{R} \times \mathbb{S}^{d}$ or
$\mathbb{R} \times \mathbb{H}^{d}$, for positive or negative
$\cc$, respectively.

When we restrict to $2+1$ dimensions the rotations commute and with
$J=-J_{12}$ we obtain the algebra
\begin{align}
  [J,P_{a}] &= \epsilon_{ab} P_{b} &  [J,D_{a}] &= \epsilon_{ab} D_{b} & [P_{a},D_{b}] &= \delta_{ab} Q \nonumber \\
  [Q,P_{a}] &=  \cc D_a & [P_a, P_b] &=  \cc \epsilon_{ab} J \,, &
\end{align}
where $\epsilon_{12}=1$. The most general invariant metric is 
\begin{align}
  \langle J, Q \rangle &= \mu  & \langle P_{a},D_{b}\rangle &=-\epsilon_{ab}\mu & \langle H, H \rangle &= \mu_{H}\nonumber
  \\
  \langle J,J \rangle &= \chi_{J} & \langle P_{a},P_{b}   \rangle &=\chi_{J} \cc \delta_{ab}& 
\end{align}
which is nondegenerate for $\mu \neq 0 \neq \mu_{H}$, i.e., we are
free to set $\chi_{J}$ to zero. The flat limit $\cc \to 0$ is well
defined on the Lie algebra and invariant metric and consequentially
also for the action. The main change is that we have the freedom to
add an additional element $\langle J,H \rangle$ to the invariant
metric which leads to
\begin{align}
  \langle J, Q \rangle &= \mu  & \langle P_{a},D_{b}\rangle &=-\epsilon_{ab}\mu & \langle H, H \rangle &= \mu_{H}\nonumber
  \\
  \langle J,J \rangle &= \chi_{J} & \langle J,H   \rangle &=\chi_{JH} \, .& 
\end{align}

\providecommand{\href}[2]{#2}\begingroup\raggedright\endgroup


\begin{thebibliography}{10}

\bibitem{Chamon:2004lew}
C.~Chamon, ``{Quantum Glassiness},''
  \href{http://dx.doi.org/10.1103/physrevlett.94.040402}{{\em Phys. Rev. Lett.}
  {\bfseries 94} no.~4, (2005) 040402},
  \href{http://arxiv.org/abs/cond-mat/0404182}{{\ttfamily
  arXiv:cond-mat/0404182}}.

\bibitem{Haah:2011drr}
J.~Haah, ``{Local stabilizer codes in three dimensions without string logical
  operators},'' \href{http://dx.doi.org/10.1103/physreva.83.042330}{{\em Phys.
  Rev. A} {\bfseries 83} no.~4, (2011) 042330},
  \href{http://arxiv.org/abs/1101.1962}{{\ttfamily arXiv:1101.1962
  [quant-ph]}}.

\bibitem{Nandkishore:2018sel}
R.~M. Nandkishore and M.~Hermele, ``{Fractons},''
  \href{http://dx.doi.org/10.1146/annurev-conmatphys-031218-013604}{{\em Ann.
  Rev. Condensed Matter Phys.} {\bfseries 10} (2019) 295--313},
  \href{http://arxiv.org/abs/1803.11196}{{\ttfamily arXiv:1803.11196
  [cond-mat.str-el]}}.

\bibitem{Pretko:2020cko}
M.~Pretko, X.~Chen, and Y.~You, ``{Fracton Phases of Matter},''
  \href{http://dx.doi.org/10.1142/S0217751X20300033}{{\em Int. J. Mod. Phys. A}
  {\bfseries 35} no.~06, (2020) 2030003},
  \href{http://arxiv.org/abs/2001.01722}{{\ttfamily arXiv:2001.01722
  [cond-mat.str-el]}}.

\bibitem{Grosvenor:2021hkn}
K.~T. Grosvenor, C.~Hoyos, F.~Pe\~na Benitez, and P.~Sur\'owka,
  ``{Space-Dependent Symmetries and Fractons},''
  \href{http://dx.doi.org/10.3389/fphy.2021.792621}{{\em Front. in Phys.}
  {\bfseries 9} (2022) 792621},
  \href{http://arxiv.org/abs/2112.00531}{{\ttfamily arXiv:2112.00531
  [hep-th]}}.

\bibitem{Brauner:2022rvf}
T.~Brauner, S.~A. Hartnoll, P.~Kovtun, H.~Liu, M.~Mezei, A.~Nicolis, R.~Penco,
  S.-H. Shao, and D.~T. Son, ``{Snowmass White Paper: Effective Field Theories
  for Condensed Matter Systems},'' in {\em {2022 Snowmass Summer Study}}.
\newblock 3, 2022.
\newblock \href{http://arxiv.org/abs/2203.10110}{{\ttfamily arXiv:2203.10110
  [hep-th]}}.

\bibitem{Cordova:2022ruw}
C.~Cordova, T.~T. Dumitrescu, K.~Intriligator, and S.-H. Shao, ``{Snowmass
  White Paper: Generalized Symmetries in Quantum Field Theory and Beyond},'' in
  {\em {2022 Snowmass Summer Study}}.
\newblock 5, 2022.
\newblock \href{http://arxiv.org/abs/2205.09545}{{\ttfamily arXiv:2205.09545
  [hep-th]}}.

\bibitem{Gromov:2017vir}
A.~Gromov, ``{Chiral Topological Elasticity and Fracton Order},''
  \href{http://dx.doi.org/10.1103/PhysRevLett.122.076403}{{\em Phys. Rev.
  Lett.} {\bfseries 122} no.~7, (2019) 076403},
  \href{http://arxiv.org/abs/1712.06600}{{\ttfamily arXiv:1712.06600
  [cond-mat.str-el]}}.

\bibitem{Slagle:2018kqf}
K.~Slagle, A.~Prem, and M.~Pretko, ``{Symmetric Tensor Gauge Theories on Curved
  Spaces},'' \href{http://dx.doi.org/10.1016/j.aop.2019.167910}{{\em Annals
  Phys.} {\bfseries 410} (2019) 167910},
  \href{http://arxiv.org/abs/1807.00827}{{\ttfamily arXiv:1807.00827
  [cond-mat.str-el]}}.

\bibitem{Pretko:2018jbi}
M.~Pretko, ``{The Fracton Gauge Principle},''
  \href{http://dx.doi.org/10.1103/PhysRevB.98.115134}{{\em Phys. Rev. B}
  {\bfseries 98} no.~11, (2018) 115134},
  \href{http://arxiv.org/abs/1807.11479}{{\ttfamily arXiv:1807.11479
  [cond-mat.str-el]}}.

\bibitem{Bidussi:2021nmp}
L.~Bidussi, J.~Hartong, E.~Have, J.~Musaeus, and S.~Prohazka, ``{Fractons,
  dipole symmetries and curved spacetime},''
  \href{http://dx.doi.org/10.21468/SciPostPhys.12.6.205}{{\em SciPost Phys.}
  {\bfseries 12} no.~6, (2022) 205},
  \href{http://arxiv.org/abs/2111.03668}{{\ttfamily arXiv:2111.03668
  [hep-th]}}.

\bibitem{Jain:2021ibh}
A.~Jain and K.~Jensen, ``{Fractons in curved space},''
  \href{http://dx.doi.org/10.21468/SciPostPhys.12.4.142}{{\em SciPost Phys.}
  {\bfseries 12} no.~4, (2022) 142},
  \href{http://arxiv.org/abs/2111.03973}{{\ttfamily arXiv:2111.03973
  [hep-th]}}.

\bibitem{Pretko:2016kxt}
M.~Pretko, ``{Subdimensional Particle Structure of Higher Rank U(1) Spin
  Liquids},'' \href{http://dx.doi.org/10.1103/PhysRevB.95.115139}{{\em Phys.
  Rev. B} {\bfseries 95} no.~11, (2017) 115139},
  \href{http://arxiv.org/abs/1604.05329}{{\ttfamily arXiv:1604.05329
  [cond-mat.str-el]}}.

\bibitem{Pretko:2016lgv}
M.~Pretko, ``{Generalized Electromagnetism of Subdimensional Particles: A Spin
  Liquid Story},'' \href{http://dx.doi.org/10.1103/PhysRevB.96.035119}{{\em
  Phys. Rev. B} {\bfseries 96} no.~3, (2017) 035119},
  \href{http://arxiv.org/abs/1606.08857}{{\ttfamily arXiv:1606.08857
  [cond-mat.str-el]}}.

\bibitem{Pena-Benitez:2021ipo}
F.~Pe\~na Benitez, ``{Fractons, symmetric gauge fields and geometry},''
  \href{http://dx.doi.org/10.1103/PhysRevResearch.5.013101}{{\em Phys. Rev.
  Res.} {\bfseries 5} no.~1, (2023) 013101},
  \href{http://arxiv.org/abs/2107.13884}{{\ttfamily arXiv:2107.13884
  [cond-mat.str-el]}}.

\bibitem{Bertolini:2023juh}
E.~Bertolini, A.~Blasi, A.~Damonte, and N.~Maggiore, ``{Gauging Fractons and
  Linearized Gravity},'' \href{http://dx.doi.org/10.3390/sym15040945}{{\em
  Symmetry} {\bfseries 15} no.~4, (2023) 945},
  \href{http://arxiv.org/abs/2304.10789}{{\ttfamily arXiv:2304.10789
  [hep-th]}}.

\bibitem{Bertolini:2023sqa}
E.~Bertolini, N.~Maggiore, and G.~Palumbo, ``{Covariant fracton gauge theory
  with boundary},'' \href{http://dx.doi.org/10.1103/PhysRevD.108.025009}{{\em
  Phys. Rev. D} {\bfseries 108} no.~2, (2023) 025009},
  \href{http://arxiv.org/abs/2306.13883}{{\ttfamily arXiv:2306.13883
  [hep-th]}}.

\bibitem{Pena-Benitez:2023aat}
F.~Pe\~na Ben\'\i{}tez and P.~Salgado-Rebolledo, ``{Fracton gauge fields from
  higher dimensional gravity},''
  \href{http://arxiv.org/abs/2310.12610}{{\ttfamily arXiv:2310.12610
  [hep-th]}}.

\bibitem{Afxonidis:2023pdq}
E.~Afxonidis, A.~Caddeo, C.~Hoyos, and D.~Musso, ``{Fracton gravity from
  spacetime dipole symmetry},''
  \href{http://dx.doi.org/10.1103/PhysRevD.109.065013}{{\em Phys. Rev. D}
  {\bfseries 109} no.~6, (2024) 065013},
  \href{http://arxiv.org/abs/2311.01818}{{\ttfamily arXiv:2311.01818
  [hep-th]}}.

\bibitem{Jain:2024ngx}
A.~Jain, ``{Fractonic solids},''
  \href{http://arxiv.org/abs/2406.07334}{{\ttfamily arXiv:2406.07334
  [hep-th]}}.

\bibitem{Hartong:2015xda}
J.~Hartong, ``{Gauging the Carroll Algebra and Ultra-Relativistic Gravity},''
  \href{http://dx.doi.org/10.1007/JHEP08(2015)069}{{\em JHEP} {\bfseries 08}
  (2015) 069},
\href{http://arxiv.org/abs/1505.05011}{{\ttfamily arXiv:1505.05011 [hep-th]}}.

\bibitem{Figueroa-OFarrill:2022mcy}
J.~Figueroa-O'Farrill, E.~Have, S.~Prohazka, and J.~Salzer, ``{The gauging
  procedure and carrollian gravity},''
  \href{http://dx.doi.org/10.1007/JHEP09(2022)243}{{\em JHEP} {\bfseries 09}
  (2022) 243}, \href{http://arxiv.org/abs/2206.14178}{{\ttfamily
  arXiv:2206.14178 [hep-th]}}.

\bibitem{Huang:2023zhp}
X.~Huang, ``{A Chern-Simons theory for dipole symmetry},''
  \href{http://dx.doi.org/10.21468/SciPostPhys.15.4.153}{{\em SciPost Phys.}
  {\bfseries 15} no.~4, (2023) 153},
  \href{http://arxiv.org/abs/2305.02492}{{\ttfamily arXiv:2305.02492
  [cond-mat.str-el]}}.

\bibitem{Marsot:2022imf}
L.~Marsot, P.~M. Zhang, M.~Chernodub, and P.~A. Horvathy, ``{Hall effects in
  Carroll dynamics},''
  \href{http://dx.doi.org/10.1016/j.physrep.2023.07.007}{{\em Phys. Rept.}
  {\bfseries 1028} (2023) 1--60},
  \href{http://arxiv.org/abs/2212.02360}{{\ttfamily arXiv:2212.02360
  [hep-th]}}.

\bibitem{Figueroa-OFarrill:2023vbj}
J.~Figueroa-O'Farrill, A.~P\'erez, and S.~Prohazka, ``{Carroll/fracton
  particles and their correspondence},''
  \href{http://dx.doi.org/10.1007/JHEP06(2023)207}{{\em JHEP} {\bfseries 06}
  (5, 2023) 207}, \href{http://arxiv.org/abs/2305.06730}{{\ttfamily
  arXiv:2305.06730 [hep-th]}}.

\bibitem{Bergshoeff:2016soe}
E.~Bergshoeff, D.~Grumiller, S.~Prohazka, and J.~Rosseel, ``{Three-dimensional
  Spin-3 Theories Based on General Kinematical Algebras},''
  \href{http://dx.doi.org/10.1007/JHEP01(2017)114}{{\em JHEP} {\bfseries 01}
  (2017) 114},
\href{http://arxiv.org/abs/1612.02277}{{\ttfamily arXiv:1612.02277 [hep-th]}}.

\bibitem{Matulich:2019cdo}
J.~Matulich, S.~Prohazka, and J.~Salzer, ``{Limits of three-dimensional gravity
  and metric kinematical Lie algebras in any dimension},''
  \href{http://dx.doi.org/10.1007/JHEP07(2019)118}{{\em JHEP} {\bfseries 07}
  (2019) 118},
\href{http://arxiv.org/abs/1903.09165}{{\ttfamily arXiv:1903.09165 [hep-th]}}.

\bibitem{Teitelboim:1983ux}
C.~Teitelboim, ``{Gravitation and Hamiltonian Structure in Two Space-Time
  Dimensions},'' \href{http://dx.doi.org/10.1016/0370-2693(83)90012-6}{{\em
  Phys. Lett. B} {\bfseries 126} (1983) 41--45}.

\bibitem{Jackiw:1984je}
R.~Jackiw, ``{Lower Dimensional Gravity},''
  \href{http://dx.doi.org/10.1016/0550-3213(85)90448-1}{{\em Nucl. Phys. B}
  {\bfseries 252} (1985) 343--356}.

\bibitem{Perez:2022kax}
A.~P\'erez and S.~Prohazka, ``{Asymptotic symmetries and soft charges of
  fractons},'' \href{http://dx.doi.org/10.1103/PhysRevD.106.044017}{{\em Phys.
  Rev. D} {\bfseries 106} no.~4, (2022) 044017},
  \href{http://arxiv.org/abs/2203.02817}{{\ttfamily arXiv:2203.02817
  [hep-th]}}.

\bibitem{Perez:2023uwt}
A.~P\'erez, S.~Prohazka, and A.~Seraj, ``{Fracton infrared triangle},''
  \href{http://dx.doi.org/10.1103/PhysRevLett.133.021603}{{\em Phys. Rev.
  Lett.} {\bfseries 133} no.~2, (2024) 021603},
  \href{http://arxiv.org/abs/2310.16683}{{\ttfamily arXiv:2310.16683
  [hep-th]}}.

\bibitem{Gromov:2018nbv}
A.~Gromov, ``{Towards classification of Fracton phases: the multipole
  algebra},'' \href{http://dx.doi.org/10.1103/PhysRevX.9.031035}{{\em Phys.
  Rev. X} {\bfseries 9} no.~3, (2019) 031035},
  \href{http://arxiv.org/abs/1812.05104}{{\ttfamily arXiv:1812.05104
  [cond-mat.str-el]}}.

\bibitem{Figueroa-OFarrill:2020gpr}
J.~Figueroa-O'Farrill, ``{On the intrinsic torsion of spacetime structures},''
  \href{http://arxiv.org/abs/2009.01948}{{\ttfamily arXiv:2009.01948
  [hep-th]}}.

\bibitem{Hansen:2020pqs}
D.~Hansen, J.~Hartong, and N.~A. Obers, ``{Non-Relativistic Gravity and its
  Coupling to Matter},'' \href{http://dx.doi.org/10.1007/JHEP06(2020)145}{{\em
  JHEP} {\bfseries 06} (2020) 145},
  \href{http://arxiv.org/abs/2001.10277}{{\ttfamily arXiv:2001.10277 [gr-qc]}}.

\bibitem{Hartong:2015zia}
J.~Hartong and N.~A. Obers, ``{Hořava-Lifshitz gravity from dynamical
  Newton-Cartan geometry},''
  \href{http://dx.doi.org/10.1007/JHEP07(2015)155}{{\em JHEP} {\bfseries 07}
  (2015) 155},
\href{http://arxiv.org/abs/1504.07461}{{\ttfamily arXiv:1504.07461 [hep-th]}}.

\bibitem{Bergshoeff:2017btm}
E.~Bergshoeff, J.~Gomis, B.~Rollier, J.~Rosseel, and T.~ter Veldhuis,
  ``{Carroll versus Galilei Gravity},''
  \href{http://dx.doi.org/10.1007/JHEP03(2017)165}{{\em JHEP} {\bfseries 03}
  (2017) 165},
\href{http://arxiv.org/abs/1701.06156}{{\ttfamily arXiv:1701.06156 [hep-th]}}.

\bibitem{Campoleoni:2021blr}
A.~Campoleoni and S.~Pekar, ``{Carrollian and Galilean conformal higher-spin
  algebras in any dimensions},''
  \href{http://arxiv.org/abs/2110.07794}{{\ttfamily arXiv:2110.07794
  [hep-th]}}.

\bibitem{Hansen:2021fxi}
D.~Hansen, N.~A. Obers, G.~Oling, and B.~T. S\o{}gaard, ``{Carroll Expansion of
  General Relativity},''
  \href{http://dx.doi.org/10.21468/SciPostPhys.13.3.055}{{\em SciPost Phys.}
  {\bfseries 13} no.~3, (2022) 055},
  \href{http://arxiv.org/abs/2112.12684}{{\ttfamily arXiv:2112.12684
  [hep-th]}}.

\bibitem{Deser:1983tn}
S.~Deser, R.~Jackiw, and G.~'t~Hooft, ``{Three-Dimensional Einstein Gravity:
  Dynamics of Flat Space},''
  \href{http://dx.doi.org/10.1016/0003-4916(84)90085-X}{{\em Annals Phys.}
  {\bfseries 152} (1984) 220}.

\bibitem{Deser:1988qn}
S.~Deser and R.~Jackiw, ``{Classical and Quantum Scattering on a Cone},''
  \href{http://dx.doi.org/10.1007/BF01466729}{{\em Commun. Math. Phys.}
  {\bfseries 118} (1988) 495}.

\bibitem{Banados:1992wn}
M.~Banados, C.~Teitelboim, and J.~Zanelli, ``{The Black hole in
  three-dimensional space-time},''
  \href{http://dx.doi.org/10.1103/PhysRevLett.69.1849}{{\em Phys.Rev.Lett.}
  {\bfseries 69} (1992) 1849--1851},
\href{http://arxiv.org/abs/hep-th/9204099}{{\ttfamily arXiv:hep-th/9204099
  [hep-th]}}.

\bibitem{Banados:1992gq}
M.~Banados, M.~Henneaux, C.~Teitelboim, and J.~Zanelli, ``{Geometry of the
  (2+1) black hole},'' \href{http://dx.doi.org/10.1103/PhysRevD.48.1506,
  10.1103/PhysRevD.88.069902}{{\em Phys.Rev.} {\bfseries D48} no.~6, (1993)
  1506--1525},
\href{http://arxiv.org/abs/gr-qc/9302012}{{\ttfamily arXiv:gr-qc/9302012
  [gr-qc]}}.

\bibitem{Kraus:2006wn}
P.~Kraus, ``{Lectures on black holes and the AdS(3) / CFT(2) correspondence},''
  {\em Lect.Notes Phys.} {\bfseries 755} (2008) 193--247,
\href{http://arxiv.org/abs/hep-th/0609074}{{\ttfamily arXiv:hep-th/0609074
  [hep-th]}}.

\bibitem{Regge:1974zd}
T.~Regge and C.~Teitelboim, ``{Role of Surface Integrals in the Hamiltonian
  Formulation of General Relativity},''
\href{http://dx.doi.org/10.1016/0003-4916(74)90404-7}{{\em Annals Phys.}
  {\bfseries 88} (1974) 286}.

\bibitem{Banados:1994tn}
M.~Banados, ``{Global charges in Chern-Simons field theory and the (2+1) black
  hole},'' \href{http://dx.doi.org/10.1103/PhysRevD.52.5816}{{\em Phys.Rev.}
  {\bfseries D52} (1996) 5816},
\href{http://arxiv.org/abs/hep-th/9405171}{{\ttfamily arXiv:hep-th/9405171
  [hep-th]}}.

\bibitem{Afshar:2016wfy}
H.~Afshar, S.~Detournay, D.~Grumiller, W.~Merbis, A.~Perez, D.~Tempo, and
  R.~Troncoso, ``{Soft Heisenberg hair on black holes in three dimensions},''
  \href{http://dx.doi.org/10.1103/PhysRevD.93.101503}{{\em Phys. Rev.}
  {\bfseries D93} no.~10, (2016) 101503},
\href{http://arxiv.org/abs/1603.04824}{{\ttfamily arXiv:1603.04824 [hep-th]}}.

\bibitem{Grumiller:2016kcp}
D.~Grumiller, A.~Perez, S.~Prohazka, D.~Tempo, and R.~Troncoso, ``{Higher Spin
  Black Holes with Soft Hair},''
  \href{http://dx.doi.org/10.1007/JHEP10(2016)119}{{\em JHEP} {\bfseries 10}
  (2016) 119},
\href{http://arxiv.org/abs/1607.05360}{{\ttfamily arXiv:1607.05360 [hep-th]}}.

\bibitem{Coussaert:1995zp}
O.~Coussaert, M.~Henneaux, and P.~van Driel, ``{The Asymptotic dynamics of
  three-dimensional Einstein gravity with a negative cosmological constant},''
  \href{http://dx.doi.org/10.1088/0264-9381/12/12/012}{{\em Class.Quant.Grav.}
  {\bfseries 12} (1995) 2961--2966},
\href{http://arxiv.org/abs/gr-qc/9506019}{{\ttfamily arXiv:gr-qc/9506019
  [gr-qc]}}.

\bibitem{Barnich:2006av}
G.~Barnich and G.~Compere, ``{Classical central extension for asymptotic
  symmetries at null infinity in three spacetime dimensions},''
  \href{http://dx.doi.org/10.1088/0264-9381/24/5/F01,
  10.1088/0264-9381/24/11/C01}{{\em Class. Quant. Grav.} {\bfseries 24} (2007)
  F15--F23},
\href{http://arxiv.org/abs/gr-qc/0610130}{{\ttfamily arXiv:gr-qc/0610130
  [gr-qc]}}.

\bibitem{Afshar:2013vka}
H.~Afshar, A.~Bagchi, R.~Fareghbal, D.~Grumiller, and J.~Rosseel, ``{Spin-3
  Gravity in Three-Dimensional Flat Space},''
  \href{http://dx.doi.org/10.1103/PhysRevLett.111.121603}{{\em Phys.Rev.Lett.}
  {\bfseries 111} no.~12, (2013) 121603},
\href{http://arxiv.org/abs/1307.4768}{{\ttfamily arXiv:1307.4768 [hep-th]}}.

\bibitem{Gonzalez:2013oaa}
H.~A. Gonzalez, J.~Matulich, M.~Pino, and R.~Troncoso, ``{Asymptotically flat
  spacetimes in three-dimensional higher spin gravity},''
  \href{http://dx.doi.org/10.1007/JHEP09(2013)016}{{\em JHEP} {\bfseries 1309}
  (2013) 016},
\href{http://arxiv.org/abs/1307.5651}{{\ttfamily arXiv:1307.5651 [hep-th]}}.

\bibitem{Ravera:2019ize}
L.~Ravera, ``{AdS Carroll Chern-Simons supergravity in 2 + 1 dimensions and its
  flat limit},'' \href{http://dx.doi.org/10.1016/j.physletb.2019.06.026}{{\em
  Phys. Lett. B} {\bfseries 795} (2019) 331--338},
  \href{http://arxiv.org/abs/1905.00766}{{\ttfamily arXiv:1905.00766
  [hep-th]}}.

\bibitem{Grumiller:2020elf}
D.~Grumiller, J.~Hartong, S.~Prohazka, and J.~Salzer, ``{Limits of JT
  gravity},'' \href{http://dx.doi.org/10.1007/JHEP02(2021)134}{{\em JHEP}
  {\bfseries 02} (2021) 134}, \href{http://arxiv.org/abs/2011.13870}{{\ttfamily
  arXiv:2011.13870 [hep-th]}}.

\bibitem{Gomis:2019nih}
J.~Gomis, A.~Kleinschmidt, J.~Palmkvist, and P.~Salgado-Rebolledo,
  ``{Newton-Hooke/Carrollian expansions of (A)dS and Chern-Simons gravity},''
  \href{http://dx.doi.org/10.1007/JHEP02(2020)009}{{\em JHEP} {\bfseries 02}
  (2020) 009}, \href{http://arxiv.org/abs/1912.07564}{{\ttfamily
  arXiv:1912.07564 [hep-th]}}.

\bibitem{Ecker:2023uwm}
F.~Ecker, D.~Grumiller, J.~Hartong, A.~P\'erez, S.~Prohazka, and R.~Troncoso,
  ``{Carroll black holes},''
  \href{http://dx.doi.org/10.21468/SciPostPhys.15.6.245}{{\em SciPost Phys.}
  {\bfseries 15} no.~6, (2023) 245},
  \href{http://arxiv.org/abs/2308.10947}{{\ttfamily arXiv:2308.10947
  [hep-th]}}.

\bibitem{Strominger:2017zoo}
A.~Strominger, {\em {Lectures on the Infrared Structure of Gravity and Gauge
  Theory}}.
\newblock Princeton University Press, 2018.
\newblock
\href{http://arxiv.org/abs/1703.05448}{{\ttfamily arXiv:1703.05448 [hep-th]}}.
\newblock

\bibitem{WZ}
X.~G. Wen and A.~Zee, ``Shift and spin vector: New topological quantum numbers
  for the hall fluids,''
  \href{http://dx.doi.org/10.1103/PhysRevLett.69.953}{{\em Phys. Rev. Lett.}
  {\bfseries 69} (Aug, 1992) 953--956}.
  \url{https://link.aps.org/doi/10.1103/PhysRevLett.69.953}.

\bibitem{Liu}
S.~Liu, A.~Vishwanath, and E.~Khalaf, ``Shift insulators: Rotation-protected
  two-dimensional topological crystalline insulators,''
  \href{http://dx.doi.org/10.1103/PhysRevX.9.031003}{{\em Phys. Rev. X}
  {\bfseries 9} (Jul, 2019) 031003}.
  \url{https://link.aps.org/doi/10.1103/PhysRevX.9.031003}.

\bibitem{You2}
Y.~You, J.~Bibo, and F.~Pollmann, ``Higher-order entanglement and many-body
  invariants for higher-order topological phases,''
  \href{http://dx.doi.org/10.1103/PhysRevResearch.2.033192}{{\em Phys. Rev.
  Res.} {\bfseries 2} (Aug, 2020) 033192}.
  \url{https://link.aps.org/doi/10.1103/PhysRevResearch.2.033192}.

\bibitem{Manjunath2}
N.~Manjunath and M.~Barkeshli, ``Crystalline gauge fields and quantized
  discrete geometric response for abelian topological phases with lattice
  symmetry,'' \href{http://dx.doi.org/10.1103/PhysRevResearch.3.013040}{{\em
  Phys. Rev. Res.} {\bfseries 3} (Jan, 2021) 013040}.
  \url{https://link.aps.org/doi/10.1103/PhysRevResearch.3.013040}.

\bibitem{May-Mann2}
J.~May-Mann and T.~L. Hughes, ``Crystalline responses for rotation-invariant
  higher-order topological insulators,''
  \href{http://dx.doi.org/10.1103/PhysRevB.106.L241113}{{\em Phys. Rev. B}
  {\bfseries 106} (Dec, 2022) L241113}.
  \url{https://link.aps.org/doi/10.1103/PhysRevB.106.L241113}.

\bibitem{Rao}
P.~Rao and B.~Bradlyn, ``Effective action approach to the filling anomaly in
  crystalline topological matter,''
  \href{http://dx.doi.org/10.1103/PhysRevB.107.195153}{{\em Phys. Rev. B}
  {\bfseries 107} (May, 2023) 195153}.
  \url{https://link.aps.org/doi/10.1103/PhysRevB.107.195153}.

\bibitem{Huang}
S.-J. Huang, C.-T. Hsieh, and J.~Yu, ``Effective field theories of topological
  crystalline insulators and topological crystals,''
  \href{http://dx.doi.org/10.1103/PhysRevB.105.045112}{{\em Phys. Rev. B}
  {\bfseries 105} (Jan, 2022) 045112}.
  \url{https://link.aps.org/doi/10.1103/PhysRevB.105.045112}.

\bibitem{Salgado-Rebolledo:2021wtf}
P.~Salgado-Rebolledo and G.~Palumbo, ``{Extended Nappi-Witten Geometry for the
  Fractional Quantum Hall Effect},''
  \href{http://dx.doi.org/10.1103/PhysRevD.103.125006}{{\em Phys. Rev. D}
  {\bfseries 103} (2021) 125006},
  \href{http://arxiv.org/abs/2102.03886}{{\ttfamily arXiv:2102.03886
  [cond-mat.mes-hall]}}.

\bibitem{Salgado-Rebolledo:2021nfj}
P.~Salgado-Rebolledo and G.~Palumbo, ``{Nonrelativistic supergeometry in the
  Moore-Read fractional quantum Hall state},''
  \href{http://dx.doi.org/10.1103/PhysRevD.106.065020}{{\em Phys. Rev. D}
  {\bfseries 106} no.~6, (2022) 065020},
  \href{http://arxiv.org/abs/2112.14339}{{\ttfamily arXiv:2112.14339
  [cond-mat.mes-hall]}}.

\bibitem{Lam:2024smz}
H.~T. Lam, J.~H. Han, and Y.~You, ``{Topological Dipole Insulator},''
  \href{http://arxiv.org/abs/2403.13880}{{\ttfamily arXiv:2403.13880
  [cond-mat.mes-hall]}}.

\bibitem{Figueroa-OFarrill:2018ilb}
J.~Figueroa-O'Farrill and S.~Prohazka, ``{Spatially isotropic homogeneous
  spacetimes},'' \href{http://dx.doi.org/10.1007/JHEP01(2019)229}{{\em JHEP}
  {\bfseries 01} (2019) 229},
\href{http://arxiv.org/abs/1809.01224}{{\ttfamily arXiv:1809.01224 [hep-th]}}.

\bibitem{Campoleoni:2022ebj}
A.~Campoleoni, M.~Henneaux, S.~Pekar, A.~P\'erez, and P.~Salgado-Rebolledo,
  ``{Magnetic Carrollian gravity from the Carroll algebra},''
  \href{http://dx.doi.org/10.1007/JHEP09(2022)127}{{\em JHEP} {\bfseries 09}
  (2022) 127}, \href{http://arxiv.org/abs/2207.14167}{{\ttfamily
  arXiv:2207.14167 [hep-th]}}.

\end{thebibliography}

\end{document}